\documentclass[preprint2]{emulateapj}
\usepackage{subfigure}
\usepackage{amsmath}
\usepackage{epstopdf}
\usepackage{rotating}
\usepackage{amsmath}

\slugcomment{accepted by AJ}

\shorttitle{BRAVA-RR}
\shortauthors{Kunder et al.}

\usepackage{graphicx}
\begin{document}

\title{The Bulge Radial Velocity Assay for RR Lyrae stars (BRAVA-RR) DR2: a Bimodal Bulge?}

\author{Andrea Kunder\altaffilmark{1},
Angeles P\'{e}rez-Villegas\altaffilmark{2},
R. Michael Rich\altaffilmark{3},
Jonathan Ogata\altaffilmark{1},
Emma Murari\altaffilmark{1},
Emilie Boren\altaffilmark{1},
Christian I. Johnson\altaffilmark{4},
David Nataf\altaffilmark{5},
Alistair Walker\altaffilmark{6},
Giuseppe Bono\altaffilmark{7,8},
Andreas Koch\altaffilmark{9},
Roberto De Propris\altaffilmark{10}
Jesper Storm\altaffilmark{11},
Jennifer Wojno\altaffilmark{5}
}
\altaffiltext{1}{Saint Martin's University, 5000 Abbey Way SE, Lacey, WA, 98503, USA}
\altaffiltext{2}{Universidade de S\~ao Paulo, IAG, Rua do Mat\~ao 1226, Cidade Universit\'aria, S\~ao Paulo 05508-900, Brazil}
\altaffiltext{3}{Department of Physics and Astronomy, University of California at Los Angeles, Los Angeles, CA 90095-1562, USA}
\altaffiltext{4}{Harvard-Smithsonian Center for Astrophysics, Cambridge, MA 02138, USA}
\altaffiltext{5}{Center for Astrophysical Sciences and Department of Physics and Astronomy, The Johns Hopkins University, Baltimore, MD 21218, USA}
\altaffiltext{6}{Cerro Tololo Inter-American Observatory, NSF’s National Optical-Infrared Astronomy Research Laboratory, Casilla 603, La Serena, Chile}
\altaffiltext{7}{Dipartimento di Fisica, Universita di Roma Tor Vergata, vi a Della Ricerca Scientifica 1, 00133, Roma, Italy}
\altaffiltext{8}{INAF, Rome Astronomical Observatory, via Frascati 33, 00040, Monte Porzio Catone, Italy}
\altaffiltext{9}{Astronomisches Rechen-Institut, Zentrum f\"{u}r Astronomie der Universit\"{a}t Heidelberg, M\"{o}nchhofstr. 12--14, 69120 Heidelberg, Germany}
\altaffiltext{10}{FINCA, University of Turku, Vesilinnantie 5, 20014, Turku, Finland}
\altaffiltext{11}{Leibniz-Institut f\"{u}r Astrophysik (AIP), An der Sternwarte 16, D-14482 Potsdam, Germany}

\begin{abstract}
Radial velocities of 2768 fundamental mode RR Lyrae stars (RRLs) toward the Southern 
Galactic bulge are presented, spanning the southern bulge 
from $\rm -8^\circ <$ $l$ $\rm < +8^\circ$ and $\rm -3^\circ<$ $b$ $\rm <-6^\circ$.
Distances derived from the pulsation properties of the RRLs are combined with 
Gaia proper motions to give constraints on the orbital motions of 1389 RRLs. 
The majority ($\sim$75\%) of the bulge RRLs have orbits consistent with these stars being 
permanently bound to $<$3.5 kpc from the Galactic Center, similar to the bar.  
However, unlike the bulge giants, the RRLs exhibit slower rotation and a higher velocity dispersion.  
The higher velocity dispersion arises almost exclusively from halo interlopers 
passing through the inner Galaxy.  
We present 82 stars with space velocities $> \sim$500~km~s$^{-1}$ and find that 
the majority of these high-velocity stars are halo interlopers;  it is unclear if a sub-sample 
of these stars with similar space velocities have a common origin.
Once the 25\% of the sample represented by halo interlopers is cleaned, we can clearly discern 
two populations of bulge RRLs in the inner Galaxy.
One population of RRLs is not as tightly bound to
the Galaxy (but is still confined to the inner $\sim$3.5~kpc), and is both spatially and kinematically
consistent with the barred bulge.  
The second population is more centrally concentrated and does not trace the bar.  
One possible interpretation is that this population was born prior to bar formation, as their 
spatial location, kinematics and pulsation properties suggest, possibly from an accretion event at high redshift.
\end{abstract}


\section{Introduction} \label{sec:intro}

The Milky Way, like most spiral galaxies, has a central bulge that 
appears to be a slightly flattened spheroid \citep{weiland94}, and 
stellar tracers, such as red clump (RC) stars and Mira variables have revealed the 
three-dimensional distribution of a bar in the Galactic bulge \citep[e.g.,][]{stanek97, 
wegg13, catchpole16}.  In our Milky Way, the bulge/bar
comprises about 15\% of the total luminosity and has a stellar mass
of $\sim$2$\times10^{10}M_\odot$ \citep{cao13, portail15a, valenti16}. 
There is much evidence of several populations coexisting within the bulge/bar, with 
these populations having different star formation histories \citep[e.g.,][]{babusiaux10, ness13, rich13, rojas19}.  

One striking example of this is the kinematically distinct rotation curves 
exhibited by the more metal-rich stars, such as the red clump (RC) stars, and 
the more metal-poor stars, such as the RR Lyrae stars (RRLs).
From radial velocities of a large sample of giants toward the Galactic bulge from the Bulge 
Radial Velocity Assay (BRAVA), the bulge was shown to be undergoing
cylindrical rotation \citep{rich07, howard09, kunder12}, and \citet{shen10} argued this
could be explained by most of the mass being in the bar.  Both the Abundances and Radial 
velocity Galactic Origins Survey \citep[ARGOS;][]{freeman13} as well as the GIRAFFE Inner Bulge 
Survey \citep[GIBS;][]{zoccali14} targeted RC stars across the bulge and confirmed cylindrical rotation of 
this population. In contrast, the Bulge Radial Velocity Assay for RR Lyrae stars, 
BRAVA-RR survey \citep[][hereafter BRR-DR1]{kunder16}, 
targeted RRLs across the inner bulge and found this population exhibited much slower rotation, as well as 
a higher velocity dispersion.  A slower rotation curve was also found for the metal-poor stars 
in \citet{arentsen20}.  Using hundreds of metal-poor inner Galaxy stars ($\rm [Fe/H]<-$1~dex), they showed 
that the signature of rotation decreases in magnitude with decreasing $\rm [Fe/H]$, 
until it disappears for the most metal-poor stars ($\rm [Fe/H]<-$2~dex). 

Both RRLs and RC stars are core helium\--burning stars widely used as 
stellar tracers of old stars to piece together the build-up of the Milky Way\citep[e.g.,][]{kunder18a}.  
However, RC stars evolve from stars with a wide range of ages ($\sim$1 Gyr to $>$10 Gyr) 
and are a more metal-rich bulge component ($\rm [Fe/H]\sim -0.5$), while RRLs 
are older ($\ge 10$ Gyr) and are on average a more metal-poor population 
\citep[$\rm Fe/H\sim -$1.2,][]{walker91}.

In contrast to ages, stellar metallicities (e.g., $\rm [Fe/H]$) can be measured using relatively well established
techniques from stellar spectra.  The metallicity distribution function (MDF) is a 
fundamental characteristic of any stellar system and stars located in the bulge have been 
differentiated by their mean $\rm [Fe/H]$ as well as by their metallicity dispersion. 
The metal-rich bulge stars are clearly part of the bulge/bar structure; they show cylindrical rotation, 
a clear vertex deviation{\footnote{The vertex deviation quantifies the misalignment  
between tangential and radial motions.  A stationary, axisymmetric disc will have no vertex deviation, 
whereas a triaxial bar necessarily introduces a vertex deviation (Binney \& Tremaine 2008).}, 
and their velocity dispersion profile decreases steeply with latitude.  
In contrast, the metal-poor stars show signs of behaving 
different, taking on a more axisymmetric spatial distribution and having lower, 
non-cylindrical rotation.  This difference in metal-rich and metal-poor stars has 
been reported from stellar abundances of the stars 
from GIBS \citep[e.g.,][]{zoccali17}, the Gaia-ESO Survey \citep[e.g.,][]{rojas17},  
from APOGEE abundances \citep[e.g.,][]{queiroz20} among others \citep[see the review by][]{barbuy18}.  
Therefore, in agreement with what is seen 
with the old RRLs in the inner Galaxy, other 
stellar populations show evidence for a metal-poor spheroid in the bulge, too.  It is still not clear, 
however, if the inner Galaxy RRLs trace the same spheroidal component as the bulge stars 
with $\rm [Fe/H] < $ 0, or if these RRLs, which are $\sim$10x more metal-poor, are yet a
different population co-existing in the bulge region of the Galaxy.

By and large, the surveys of the bulge mentioned above have lacked 
accurate proper motions, allowing an exploration of the bulge in only one velocity component.
Samples of stars with measured proper motions have also indicated that the bulge harbors
diverse stellar components.  \citet{clarkson18} showed that the velocity ellipsoid of 
metal-rich stars is different from the subsolar stars, and \citet{soto07}
finds that the metal-rich bulge stars have a vertex deviation in the radial versus transverse 
velocity, consistent with the bar supporting orbits.  However, using proper motions 
of the entire Galactic bulge region from VIRAC and Gaia, \citet{clarke19} show that 
the inner bulge stars have correlated proper motions and no clear separation of different 
bulge components were detected.

Given this stellar potpourri, the speculation has arisen that the more metal-poor stars in the 
direction of the bulge are not actually of the bulge, but are part of the halo, as the halo of 
the Milky Way obviously is also present in the inner Galaxy \citep{johnson12, kunder15, koch16}.  
These stars would be inner halo stars on orbits that make them pass through the central 
regions.  They could also be the inner-extension of the halo confined to 
the inner Galaxy \citep{perezvillegas17a} $i.e.,$ these stars are metal-poor halo stars 
with orbits that confine them to the inner $\sim$5~kpc of the Galaxy.  The majority of these 
stars would not follow the bar/X-shaped structure and would show a slow mean radial velocity 
with a velocity dispersion of $\sim$120~km~s$^{-1}$.  
The dearth of 6D phase space information for bulge stars has made it difficult 
to explore the dynamics of large sample of
bulge stars, however.  For the small stellar samples with 6D information, it is seen that 
at least some of the metal-poor bulge stars are confined to the bulge \citep{howes15}.

Here, we refine the radial velocity of the inner Galaxy RRL stars presented in BRR-DR1
and present new radial velocities for RRLs in six bulge fields located in the outer bulge 
in \S2. 
Our sample consists of 2768 RRLs, 2.5 times larger than in BRR-DR1, 
and the radial velocities are presented in \S3.  The fastest moving stars are 
discussed in \S4.  In \S5, we focus on the $\sim$50\% of these stars with Gaia astrometry (Gaia DR2).  
We use the 3D velocities (and orbits) to study halo or bulge membership of our inner Galaxy RRL sample.
Our findings, that there are two different spatial distributions of RRLs in the bulge, are put into context in \S6.

\section{Data}

\begin{figure}
\centering
\mbox{\subfigure{\includegraphics[height=8.0cm, width=6.0cm]{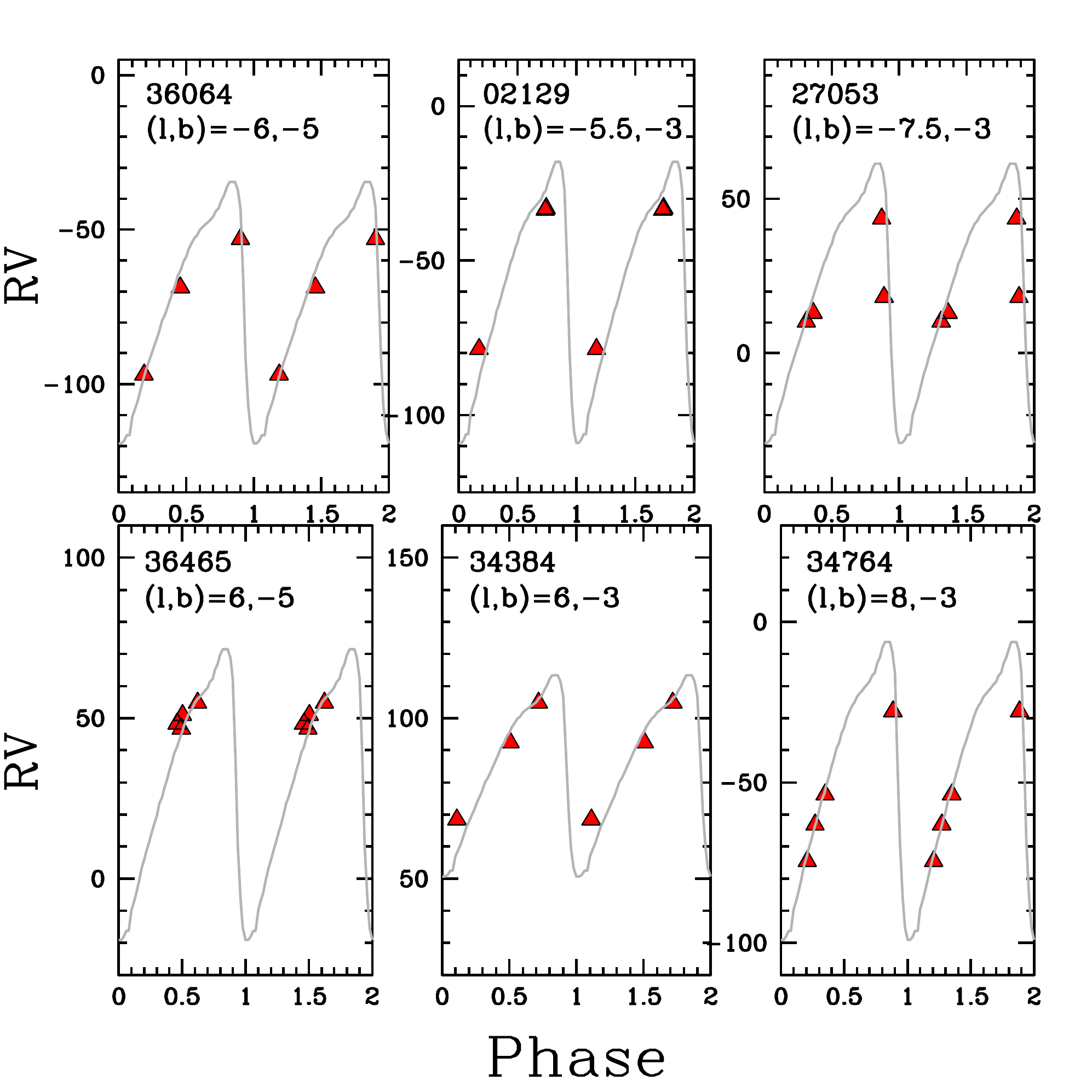}}}
{\subfigure{\includegraphics[height=7.7cm, width=2.5cm]{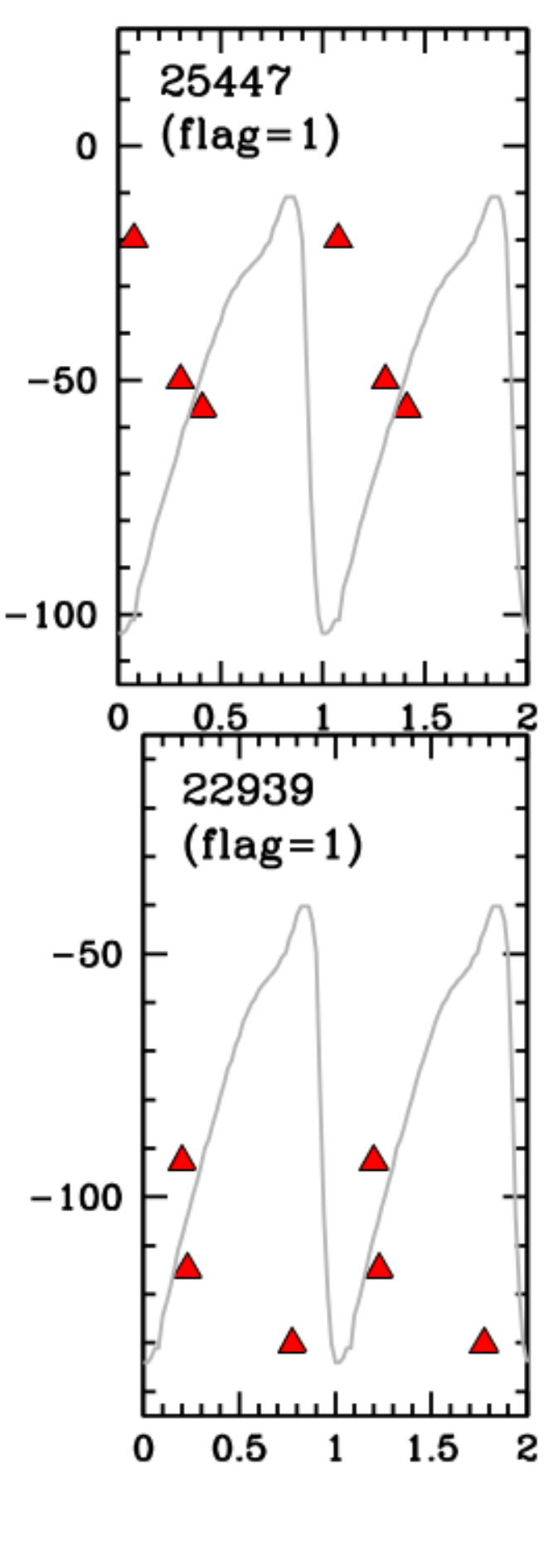}}}
\caption{A representative selection of light curves for the inner Galaxy RRL stars presented here.  The radial 
velocity template used is over-plotted \citep{liu91}.  Stars with {\tt flag=1} indicate RRLs with 
scatter about their radial velocity curves.}
\label{rvcurve}
\end{figure}


\begin{figure*}
\centering
\subfigure{
\includegraphics[height=6.0cm]{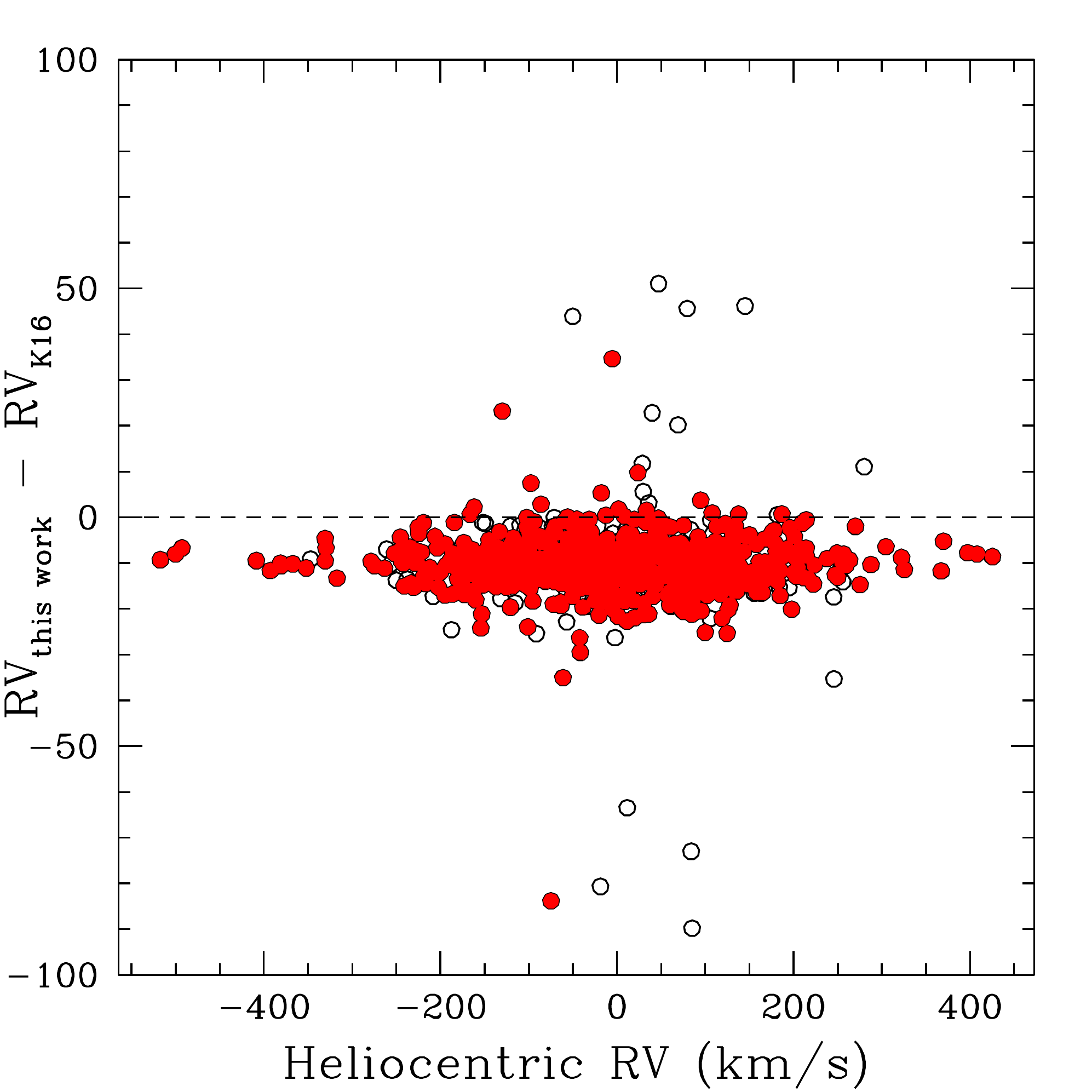}
}
\subfigure{
\includegraphics[width=6.0cm]{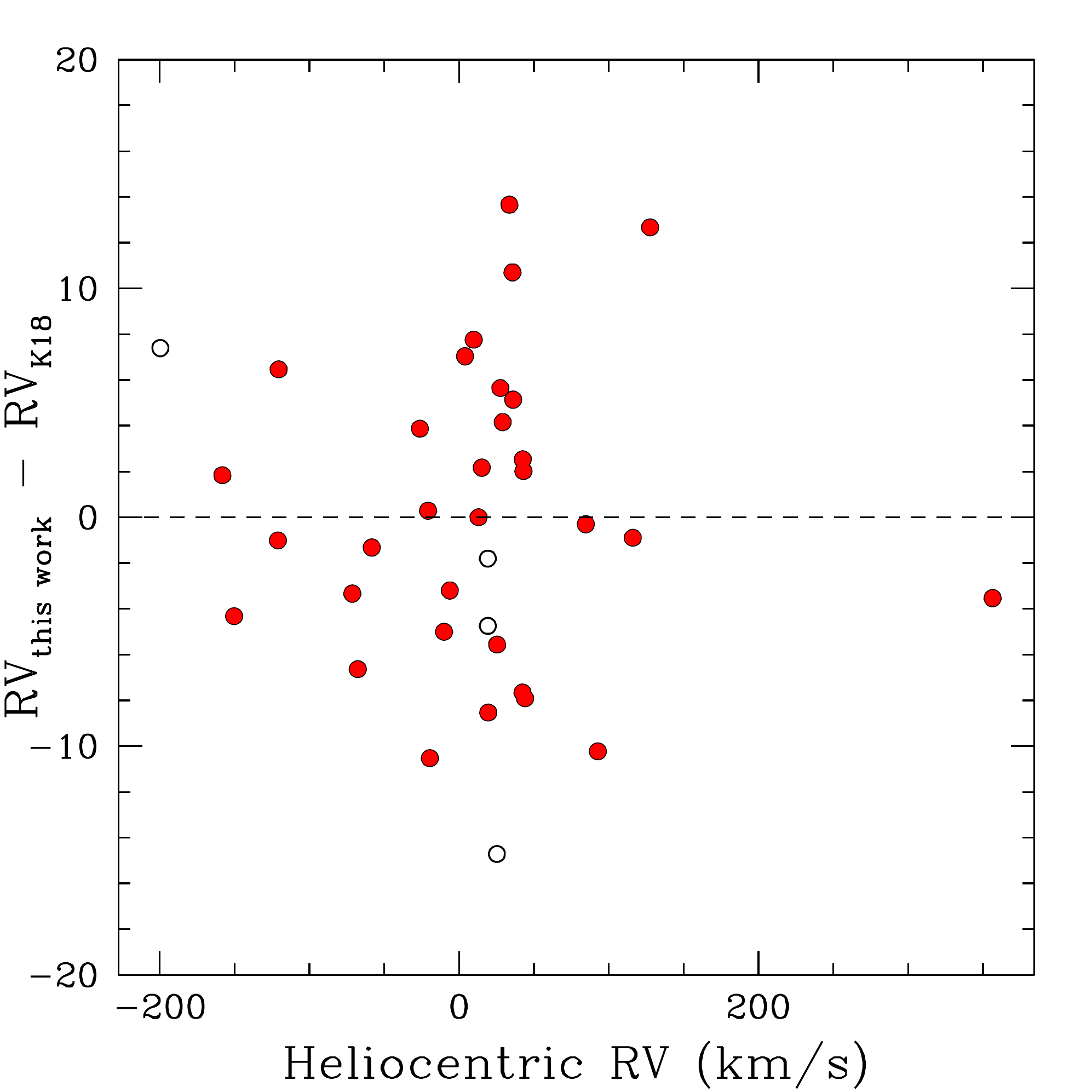}
}
\caption{{\it Left:} The difference in radial velocities is shown between BRR-DR1 and those presented here.  The filled (red) circles indicate those with high quality velocities, $i.e.$, {\tt flag=0}, whereas those with open circles indicate those with more uncertain radial velocities, $i.e.$, {\tt flag=1} 
{\it Right:} The difference in radial velocities is shown between those from \citet{kunder18b} and those presented here.}
\label{comps}
\end{figure*}

\subsection{Radial velocities}
The OGLE-IV catalog of bulge RRLs \citep{pietrukowicz15} was used to select the RRLs.
As in BRR-DR1, the wavelength covered was $\sim$8300\AA~to 8800\AA~at a resolution 
of R$\sim$10,000 using the AAOmega multi-fiber spectrograph on 
the Anglo-Australian Telescope (AAT).  
Exposure times were between 0.33 to 3 hours, and in general we collected between
1 - 4 epochs for each RRL, the multiple observations required to measure systemic velocities for each star.  
Observations were taken from June 16 - June 21, 2017\footnote{NOAO PropID 2017A-0196
and OPTICON 51}.  Reductions were carried out in an identical manner as BRR-DR1, and radial 
velocities were found using iraf's {\tt xcsao} routine.  The {\tt xcsao} cross-correlatation was carried 
out using spectra taken of the APOGEE star 2M18264551-1747096 and adopting the velocity of this star 
to be $-$69.2~km~s$^{-1}$ \citep{alam15}.  The spectrum of this star was acquired using 
the same set-up with AAOmega that was used with our RRLs.  The individual uncertainties from {\tt xcsao} 
typically ranged from $\sim$1~km~s$^{-1}$ to $\sim$4~km~s$^{-1}$.   

The radial velocity measurements were phased by the stars' known OGLE period, and 
over-plotted with the radial velocity template from \citet{liu91}.  The template is scaled using
\begin{equation}
A_{rv} = \frac{40.5\times V_{amp} + 42.7}{1.37},
\end{equation}
where $V_{amp}$ is the $V$-band amplitude, 
to account for the correlation between the amplitudes of velocity 
curves and light curve \citep{liu91}.  To be consistent with BRR-DR1, we adopt
$p$= 1.37, the so-called ``projection factor", to relate our observed radial velocities to
the pulsation radial velocities \citep[see also][]{sesar12}.  Because OGLE samples 
the $I$-band much more frequently than the $V$-band, we used the following relation 
for the $V$-amplitude:
\begin{equation}
V_{amp} = I_{amp}\times1.6
\end{equation}
\citep[see e.g., Table~3 in][]{kunder13}.  

In BRR-DR1, we found that the uncertainty in the center-of-mass radial velocity measurement due 
to having a small (between 1-4) number of epochs to be between 2-5~km~s$^{-1}$. 
Adding in quadrature both the individual uncertainty on each velocity measurement as reported 
from  {\tt xcsao} as well as that from having a small number of epochs for 
center-of-mass radial velocity determination, the total center-of-mass radial velocity uncertainties
are $\sim$5-10 km~s$^{-1}$.

Although we made every effort to observe RRLs that were free of a companion 
within a 2 arc second radius, we did notice evidence of blending in a small fraction 
of our RRL stellar spectra ($\sim$1\%).  For some spectra, not one, but two distinct dips in intensity 
around the calcium triplet(CaT) was seen.  These two distinct CaT lines represent the CaT 
lines of two different stars: the RRL as well as a close neighbor.  The most likely reason for this blending 
is that the typical seeing during our observing runs was between 1.5 - 2.5 arc seconds.  
We discarded the spectra with two distinct CaT lines, as we could not be sure which CaT line belonged to 
the RRL and which one belonged to the neighbor.  There may be other stars that have blends 
but that we were not able to identify due to low SNR, a small difference in radial velocity between the RRL and the 
neighboring star and/or too many spectral lines near the CaT lines from e.g., shock waves \citep{chadid08, yang14}
or the star having a hot temperature where Hydrogen Paschen lines dominate as compared to the CaT signature. 

For 98\% of our sample, we had more than one epoch of observations.  Therefore, we 
visually checked each radial velocity curve and made sure it matched that from \citet{liu91} using the
zero-point in phase fixed using the OGLE-IV time of maximum brightness. 
Stars with large scatter in their radial velocity curves are flagged with {\tt flag=1}, where large 
scatter means that at least one observation deviates by $\sim$20~km~s$^{-1}$ or more from the 
\citet{liu91} template.  Two typical stars with {\tt flag=1} are shown in Figure~\ref{rvcurve}.

Some of the stars with scatter may be attributed to stellar blending that went undetected 
by our visual examinations of stellar spectra.  Blends are less likely to be detected if a particular 
observation is taken when an RRL is at a high temperature during its pulsation cycle, 
as this is where the Hydrogen Paschen lines become 
significant and the CaT lines are less prominent.  We do not believe that stars with {\tt flag=0} 
are blends, as otherwise the individual radial velocities would not follow the \citet{liu91} 
template (see Figure~\ref{rvcurve}).  
Note that the phase of the observation is fixed using the times of minimum brightness 
given by OGLE, and the \citet{liu91} template is shifted only up or down to fit the individual 
radial velocity measurements.  
We can not be certain that stars with one epoch of observations are not affected by 
undetected blends.  

Figure~\ref{comps} shows a comparison between the radial velocities presented here and 
those from both BRR-DR1 and from \citet{kunder18b}.  A striking $\sim$10~km~s$^{-1}$ offset is 
found from the radial velocities here as compared to BRR-DR1, and no meaningful 
offset is found from the radial velocities here as compared to \citet{kunder18b}.  The difference in the 
BRR-DR1 velocities stems from the standard star used for the radial velocity zero-point.  In contrast to 
BRR-DR1, where a star from the BRAVA survey was used as a radial 
velocity standard, here we set the radial velocity zero point 
to the APOGEE star 2M18264551-1747096.  The radial velocity of this star is better determined 
than the one used by BRR-DR1, and therefore 
the velocities presented here should supersede those from BRR-DR1.  In \citet{kunder18b}, the center 
of the three CaT lines were measured directly by hand, so the agreement between these radial velocities 
and the ones presented here indicates that our zero-point is much improved from BRR-DR1.  

Examples of RRL radial velocity curves are shown in Figure~\ref{rvcurve}, where we show 
those stars with the largest space velocities, which are discussed further in \S\ref{sec:emilie}.
All radial velocity measurements are presented in Table~\ref{lcpars}.
Table~\ref{lcpars} gives the OGLE-ID (1), 
the RA (2) and Dec (3) as provided by OGLE, 
the star's time-average velocity (4), 
the number of epochs used for the star's time-average velocity (5), 
the radial velocity flag (where stars with well-fit radial velocity curves have {\tt flag=0} and stars with 
large scatter around their radial velocity curves are flagged with {\tt flag=1}) (6), 
the period of the star (7), 
$V$-band magnitude (8),
the $I$-band magnitude (9) 
and the $I$-band amplitude (10) as calculated by OGLE, 
and lastly the distance (see \S\ref{sec:dist}) adopted for the orbital integration (11).

\subsection{Proper motions}
\label{sec:pm}
With the Gaia DR2 data release (Gaia Collaboration et al. 2018 ), we now have in hand 
millions of proper motions.  These have (for example) the precision to separate bulge 
RRLs from those RRLs residing in the Sagittarius dwarf galaxy (Sgr) 
\citep[e.g.,][]{kunder19} that otherwise complicate the study of the bulge.

Proper motions of the RRLs are obtained from cross-matching with the Gaia DR2 catalog.  To ensure 
that the cross-match for each RRL is correct in the crowded bulge region of the sky, only the 
matches with the Gaia flag {\tt phot{\_}variable{\_}flag = VARIABLE} were kept.  We further limited 
our cross-matches to those stars with the recommended astrometric quality indicators such as the 
unit weight error (UWE) and the renormalised unit weight error (RUWE), as described by \citet{lindegren18}.  
Specifically, only stars with a RUWE $<$ 1.4 were kept, where the RUWE of a star was calculated using
\begin{equation}
RUWE = \frac{UWE}{\mu_{0}(G,C)}.
\end{equation}
Here, $\mu_{0}(G,C)$ is a normalization factor obtained from interpolating 
the $phot{\_}g{\_}mean{\_}mag$ (G) and $rp - bp$ color of the stars on the table on the 
ESA $Gaia$ DR2 known issues page. The UWE can be calculated using the equation
\begin{equation}
UWE = \sqrt{\frac{\chi^2}{(N-5)}},
\end{equation}
where $\chi^2$ is given by the $astrometric{\_}chi2{\_}al$ parameter and N is given by 
the $astrometric{\_}n{\_}good{\_}obs{\_}al$ parameter. 
Our final sample of RRLs with Gaia DR2 astrometry contains 1389 stars out of the 
2768 RRLs presented here.

\subsection{Distances}
\label{sec:dist}
\begin{figure*}
\centering
\mbox{\subfigure{\includegraphics[height=6.5cm]{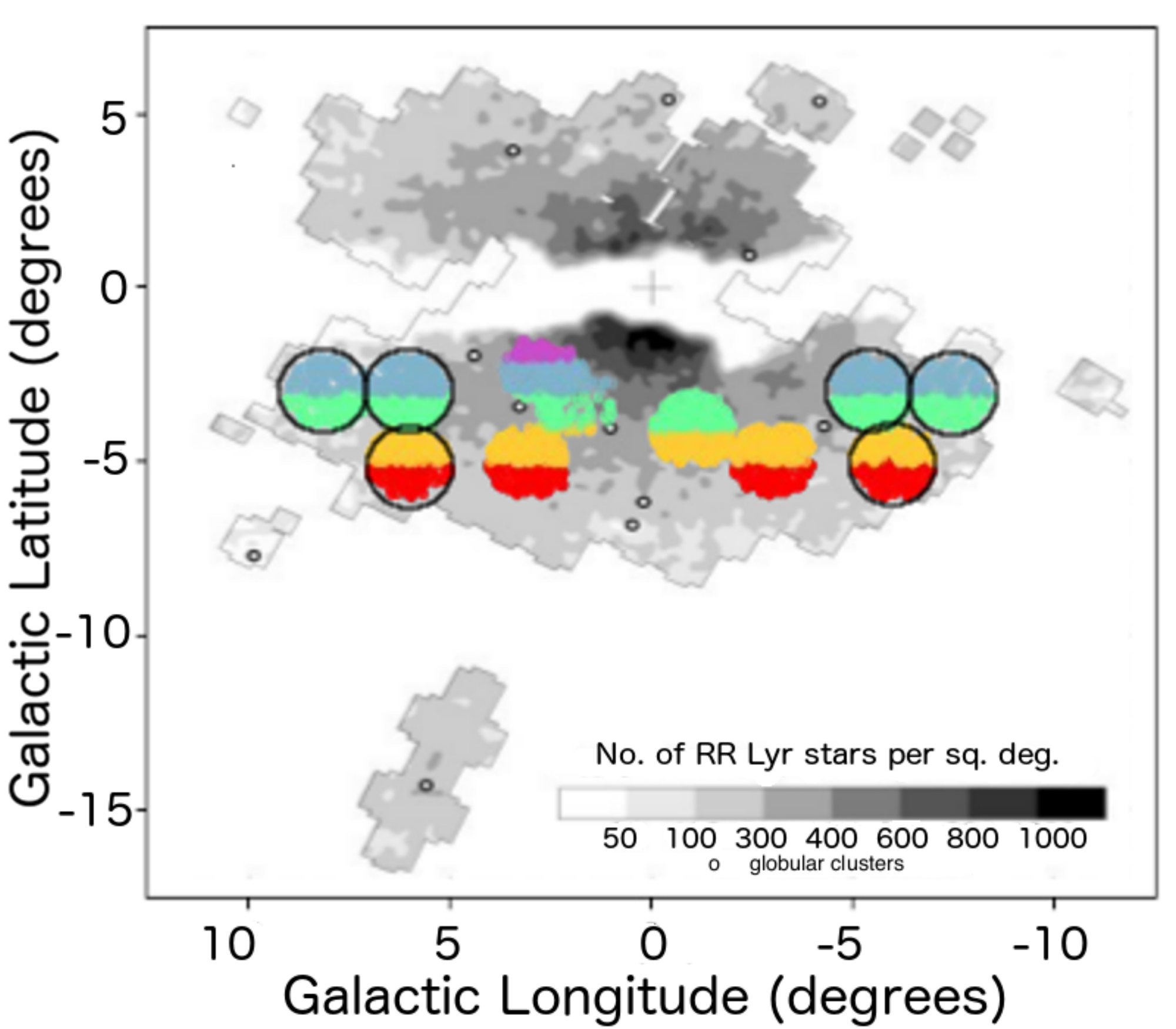}}
          \subfigure{\includegraphics[width=9.0cm]{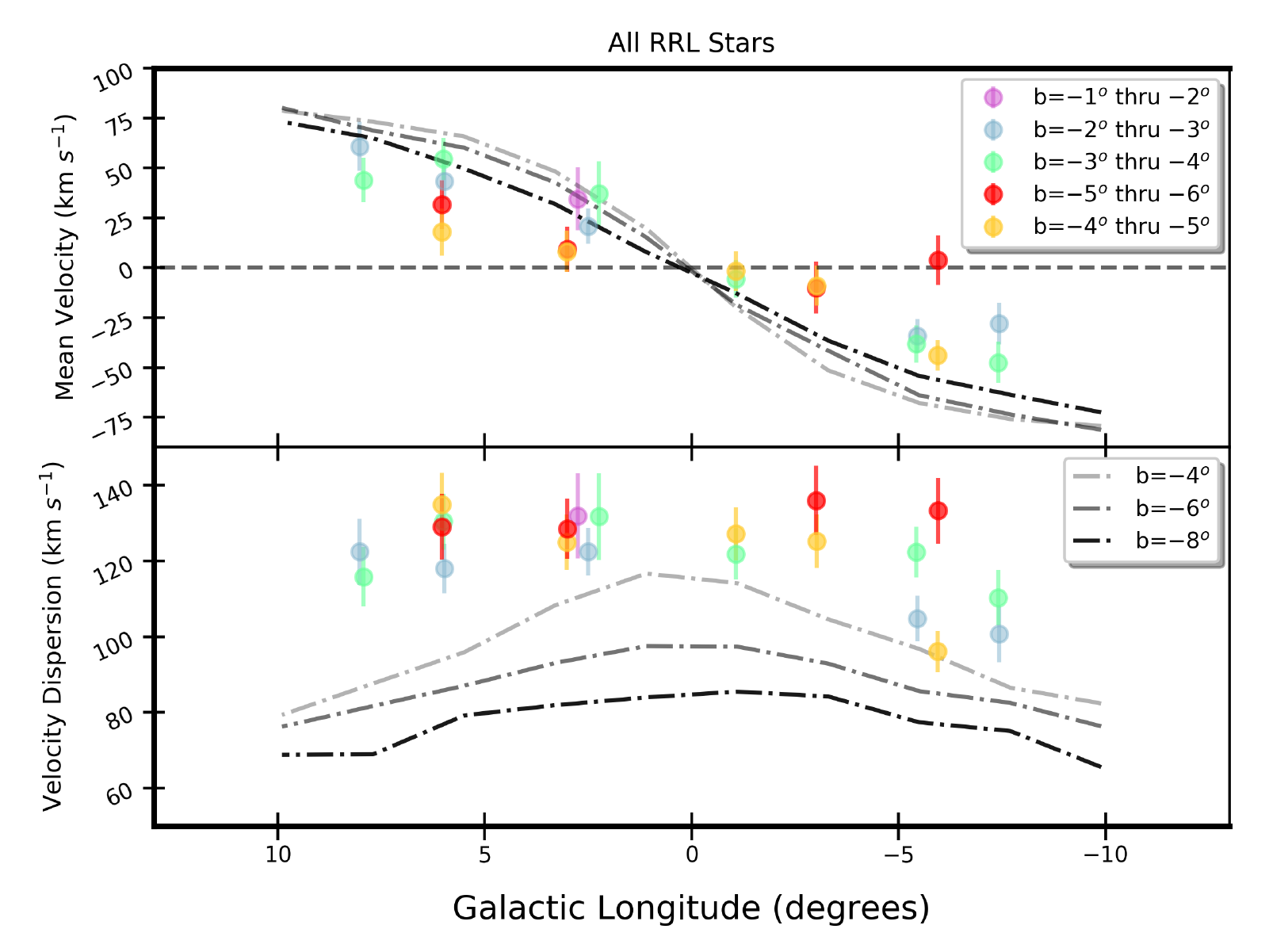}}}
\caption{{\it Left:} The spatial location of the OGLE RRLs in the Galactic bulge, where our 
observed RRLs are color coded to 
designate the strips of latitude they are separated into, with which to obtain their rotation 
curve (see right panels).  New fields are shown by the black outline.  {\it Right:} 
The longitude versus mean galactocentric velocity, GRV, (top) and 
the longitude versus mean standard deviation of GRV (bottom) for the RRLs presented here 
compared to that of the BRAVA giants at 
$b$ = $-$4$^\circ$, $-$6$^\circ$, and $-$8$^\circ$ strips \citep{kunder12}.  
The colors correspond to the stars with spatial positioins as indicated in the left panel (see also legend).
Lines from the bar model from \citet{shen10} have been over-plotted for 
$b$=$-$4$^\circ$ (grey), $-$6$^\circ$ (dark grey), $-$8$^\circ$ (black).
}
\label{fields}
\end{figure*}
One advantage of using RRLs as stellar tracers is that they are relatively well-understood standard candles.  
Distances for each RRL are calculated using the identical procedure as 
in \citet{pietrukowicz15}.  First, the photometric metallicities of the stars were found using the following relation from \citet{smolec05}:
\begin{equation}
\rm [Fe/H] = -3.142 - 4.902P + 0.824(\phi_{31}+\pi))
\end{equation}
where P is the pulsation periods of the stars and $\phi_{31}$ is 
a Fourier phase combination for the cosine decomposition for the $I$-band 
light curves of the stars.  The photometric metallicities indicate a sharply peaked 
RRL metallicity distribution at $\rm [Fe/H]$ = $-$1.02~dex, with a dispersion of 0.25~dex 
\citep{pietrukowicz12, pietrukowicz15}.
These photometric metallicities were then used to find their absolute magnitudes, $M_V$ and $M_I$, using theoretical relations from \citet{catelan04}:
\begin{equation}
M_V = -2.288 - 0.882~\rm{log(Z)} + 0.108~log(Z)^2
\end{equation}
\begin{equation}
M_I = -0.471 - 1.132~\rm{log(P)} + 0.205~\rm{log(Z)}
\end{equation}
where 
\begin{equation}
\rm{log(Z)} = [Fe/H] - 1.765.
\end{equation}
Because we use $\rm [Fe/H]$ instead of $\rm [M/H]$, we are not taking into account any effect of 
an enhancement in $\alpha$-capture elements with respect to a solar-scaled mixture.  This is because 
there are no 
studies that have been carried concerning the $\alpha$-abundances of bulge field RRLs (note that the
study by Hansen et~al. 2016 concerned one RRL in the direction of the bulge that is thought to be a 
halo interloper and not a bonafide bulge field RRL star).  Also, even assuming the bulge RRL are 
considerably enhanced with $\rm [\alpha/Fe]$=0.6, this would change our log $Z$ by $\sim$0.5, 
resulting in a distance change of $\sim$2\%.  A more modest enhancement of 
$\rm [\alpha/Fe]$=0.2 would change our log $Z$ by $\sim$0.15, 
resulting in a distance change of $\sim$0.5\%.  

The reddening along the line of sight of the star is obtained using the following relation from \citet{nataf13}:
\begin{equation}
A_I = 0.7465[(m_V - m_I) - (M_V - M_I)] + 1.37E(J-K)
\end{equation}
where $m_V$ and $m_I$ are the OGLE mean magnitudes of the RRL in the $V$ and $I$ passbands, respectively, 
and $\rm E(J-K)$ is the reddening from \citet{gonzalez12}. 
Lastly, the distance can be found using
\begin{equation}
d = 10^{0.2 (m_I - M_I - A_I + 5)}
\end{equation}
Assuming $\rm [Fe/H] = -1$ for all RRLs changes the individual distances by $\sim$1.5\%.
Assuming $\rm [Fe/H] = 0$ for all RRLs changes the individual distances by $\sim$5\%, and the 
distances to smaller/closer values.  Assuming  $\rm [Fe/H] = -2$ for all RRLs changes the individual 
distances by $\sim$8\%, shifting the distances to further/larger values.  The 10\% uncertainty in 
distance, as assumed throughout this paper, is not dominated by the uncertainties in metallicity, but 
rather the uncertainties in reddening and in the theoretical $M_V$ and $M_I$ relations.
In this paper, the RRL distances are used mainly to compare the RRLs to each other ($e.g.,$ those 
RRLs with distances closest to the Galactic center and those with distances farther from the Galactic center).
Relative distances are not as independent on the exact zero-point used in $M_V$.

From the RRLs studied here, we find a median distance of 8.48~kpc, with a standard deviation of 1.6~kpc.
It has been shown by \citet{pietrukowicz15} that when 
projecting the individual distances using this method onto the Galactic plane (by cos $b$) and scaling the distribution by d$^{-2}$ to compensate for the ``cone effect" ($i.e.$, where more objects are seen at larger distances inside a solid angle), a distance to the Galactic center of $R_0 =$ 8.27~kpc$\pm$0.008(stat)$\pm$0.40(sys) kpc is obtained.  They also show that using other absolute magnitude relations, e.g., \citet{marconi15,bono07}, do not affect the distance determinations significantly.

\section{Orbits}
\label{sec:orbit}
For the orbital analysis,  we employed a non-axisymmetric model for the Galactic gravitational potential. 
The model has an axisymmetric background made by an exponential disk built from the superposition of 
three Miyamoto-Nagai potentials \citep{miyamoto75} following the recipe by \citet{smith15}, 
and a Navarro-Frenk-White density profile \citep{navarro97} to model the dark matter halo, which 
has a circular velocity $V_0 = 241$~km~s$^{-1}$ at $R_0 = 8.2$~kpc \citep{blandhawthorn16}.  
A triaxial Ferrer’s bar potential is superimposed on the axisymmetric background. The total bar mass 
is $1.2 \times 10^{10}~M_\odot$, an angle of $25^\circ$ with the Sun-major axis of the bar, a pattern 
speed of the bar of  $\Omega_b=40$~km~s$^{-1}$~kpc${-1}$ \citep{portail17, perezvillegas17b}, 
and a major axis extension of 3.5~kpc. 

As the bar mass is still under debate, the adopted bar mass ($1.2 \times 10^{10}~M_\odot$) 
comes from the dynamical models 
by \citep{portail15b}.  These models reproduce the kinematics from the BRAVA survey \citep{rich07, kunder12}. 
The angle of 25$^\circ$ with the Sun-major axis of the bar is based on OGLE-II red clump giant stars 
\citep{rattenbury07}, this value is in good agreement with the measurement obtained from VVV red clump 
stars \citep{wegg13}.
The viral mass of our DM halo is $\rm 3.41x10^{12}~M_\odot$, which is the mass contained 
within 302 kpc \citep{perezvillegas20}.  This corresponds to a density of 200 times the 
critical density, in agreement with \citet{abadi09} and \citet{irrgang13}.

Of course, an orbital analysis is not without limitations.  The steady potential we adopted includes a 
triaxial bar, an exponential disk, and a spherical dark matter halo. 
As a steady model does not evolve with time, we are not able to study secular evolution in the 
Galaxy and there are no migratory processes that modify the potential.
We also do not consider the dynamical contribution due to spiral arms because 
within 4~kpc from the Galactic center, their effects are negligible.  This is because the mass of 
the spiral arms ($\sim$5\% of the disk mass) is small compared with the bar mass 
($\sim$15-20\% of the disk mass).  Also, it has been shown that the spiral arms have important 
dynamical effects between 3-10~kpc, the radii they dominate \citep[e.g.,][]{allen08, antoja11}.

Some fraction of the inner bulge could be due to accretion events of satellite galaxies. If the 
accretions happened before bar formation, the distribution and dynamics of the accreted material 
will be gravitationally affected by the bar formation and could be trapped by the bar to give it orbital support. 
With an existing bar component, stars that reach the bulge/bar region because of a disrupting satellite galaxy 
due to dynamical friction, will still keep their dynamical memory from their origin, but that memory could be 
deleted by secular evolution with time.  With the model we use, we cannot take 
such processes into account.

The greatest limitation of our steady potential is that we cannot follow the secular evolution of 
structure formation within the Milky Way.  Therefore, there is little difference 
in the orbital parameters when integrating forward or backward in time.  
In an attempt to mitigate the effects of a changing potential, we 
integrated the orbits for $\sim$5~Gyr.  Cosmological simulations of Milky-Way-like galaxies \citep[e.g.,][]{buck18} 
and analysis of field stars \citep[e.g.,][]{bovy19} show that the Galactic bar likely formed 
about $8 \pm 2$ Gyr ago.  This is when the Galactic potential would have had the most significant change, 
and our orbital integrations avoid this time period.  
Some Galactic regions may have had chaotic behavior even after bar formation.  In these regions, 
forward and backward integration would not be same.  This is beyond the scope of our study.  

The velocity components of the Sun with respect to the local standard of rest are 
$(U, V, W )_\odot$ = (11.1, 12.24, 7.25) km s$^{-1}$ \citep{schoenrich10}, where 
U, V, W, are positive in the direction of the Galactic center, Galactic rotation, and North Galactic Pole, respectively. 
To estimate the effect of the uncertainties associated to the observational data, we generate a set of 
100 initial conditions for each RRL, using a Monte Carlo method, where we take into account an uncertainty 
of 10\% for the distance determination, an uncertainty of 10 km s$^{-1}$ for the heliocentric radial velocity, and an 
uncertainty of absolute proper motion in both components that includes the average of the systematic error 
of $0.035$ mas yr$^{-1}$ reported by the Gaia Collaboration et al. (2018).  
We integrate these initial conditions forward for 5 Gyr. 

For each orbit, we calculate the perigalactic distance, apogalactic distance, the maximum vertical excursion 
from the Galactic plane $|z|_{max}$, and the eccentricity defined by $e= (apo-peri)/(apo + peri)$.

\section{Rotation Curve}
BRR-DR1 was the first to show that the inner Galaxy RRLs do not rotate like the bulge giants 
that trace out the bar/bulge, and that they also exhibit larger velocity dispersions.  Because of the 
old age of RRLs, they attributed this to the inner Galaxy RRL sample as having formed before 
the bar/bulge.  Since then, the rotation curve of hundreds of metal-poor stars located 
toward the direction of the bulge confirmed that metal-poor stars have slower rotation and larger 
velocity dispersions than the bar/bulge giants \citep{arentsen20}.

Figure~\ref{rvcurve} shows the spatial location of the studied RRLs; the new BRAVA-RR 
fields extend to a Galactic longitude of 5$^{\circ} < |l| > 9^{\circ}$.  The rotation curve of these 
stars is also shown in Figure~\ref{rvcurve}.  Here, the heliocentric velocities are corrected to 
galactocentric velocity, GRV, taking into the 
solar reflex motion by
\begin{multline}
GRV = V_{HC} + 220~\rm sin({\it l})cos({\it b})+\\
16.5[\rm sin({\it b})sin(25) + cos({\it b})cos(25)cos({\it l}-53)],
\end{multline}
where $V_{HC}$ is the mean RRL heliocentric velocity.
The bulge model from \citet{shen10} is over-plotted on the RRL rotation curve.  
The \citet{shen10} model shows what it is expected for a bulge being formed from the disk 
and undergoing cylindrical rotation, and is consistent with observations of bulge 
giants \citep{kunder12, ness16, zoccali14}.

In agreement with what was found in BRR-DR1, the inner Galaxy RRLs are not rotating 
in the same manner as the giants.  The cylindrical rotation seen in bar/bulge stars, in a sense that 
the mean velocity does not change as a function of Galactic latitude, is absent in the bulge RRLs.  Instead,
the mean velocity decreases as a function of latitude, which would be expected if the RRLs are 
rotating in a spherical-like manner, or alternatively, if the rotation curve is being contaminated with 
stars in the thin/thick disk or halo with increasing distance from the plane.  In \S\ref{sec:2pops} we
use proper motions to clean our sample from interlopers to investigate the rotation curve of a purer bulge 
RRL sample.

\begin{figure*}
\centering
\mbox{\subfigure{\includegraphics[height=6.0cm]{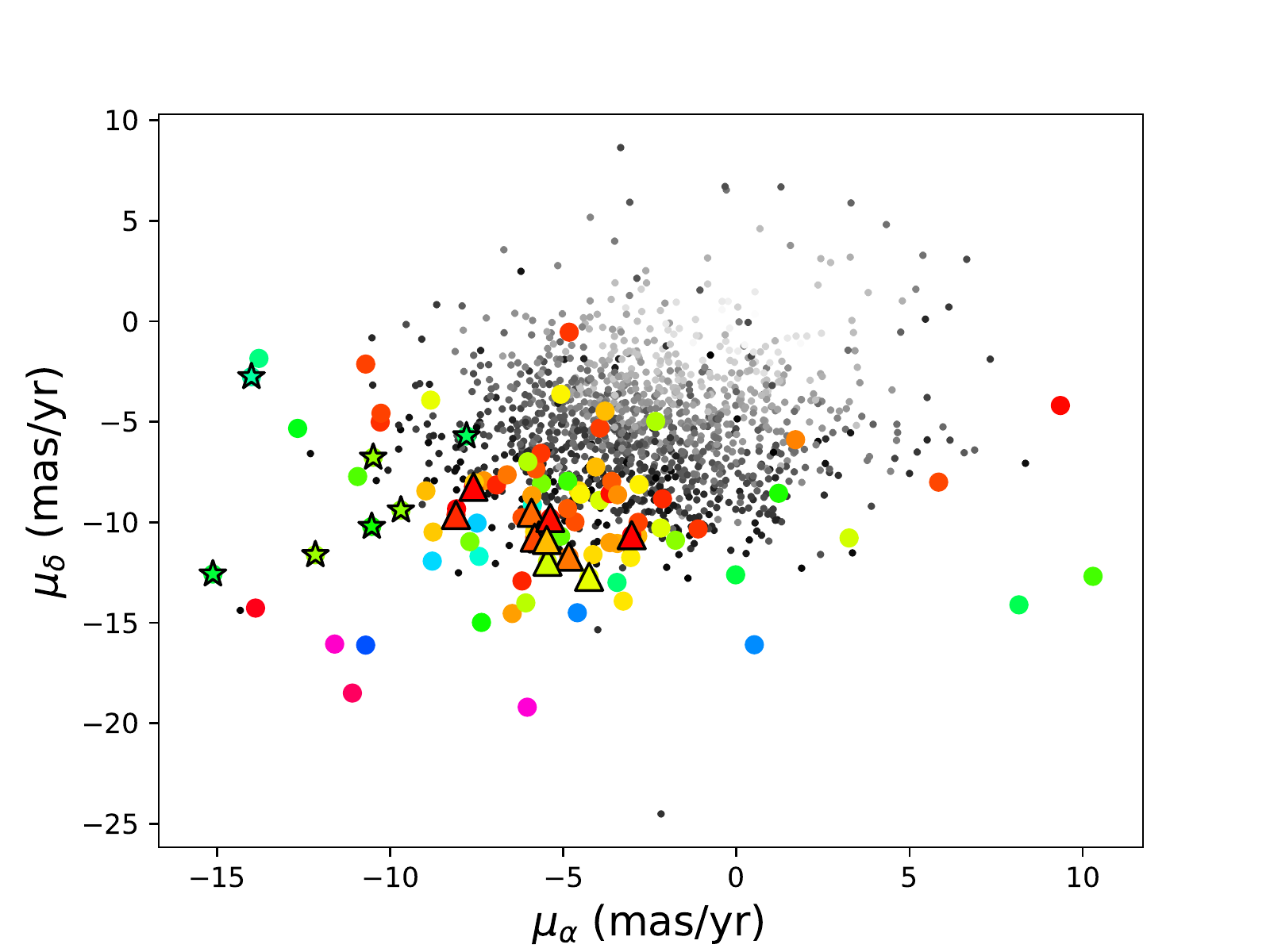}}
          \subfigure{\includegraphics[height=6.0cm]{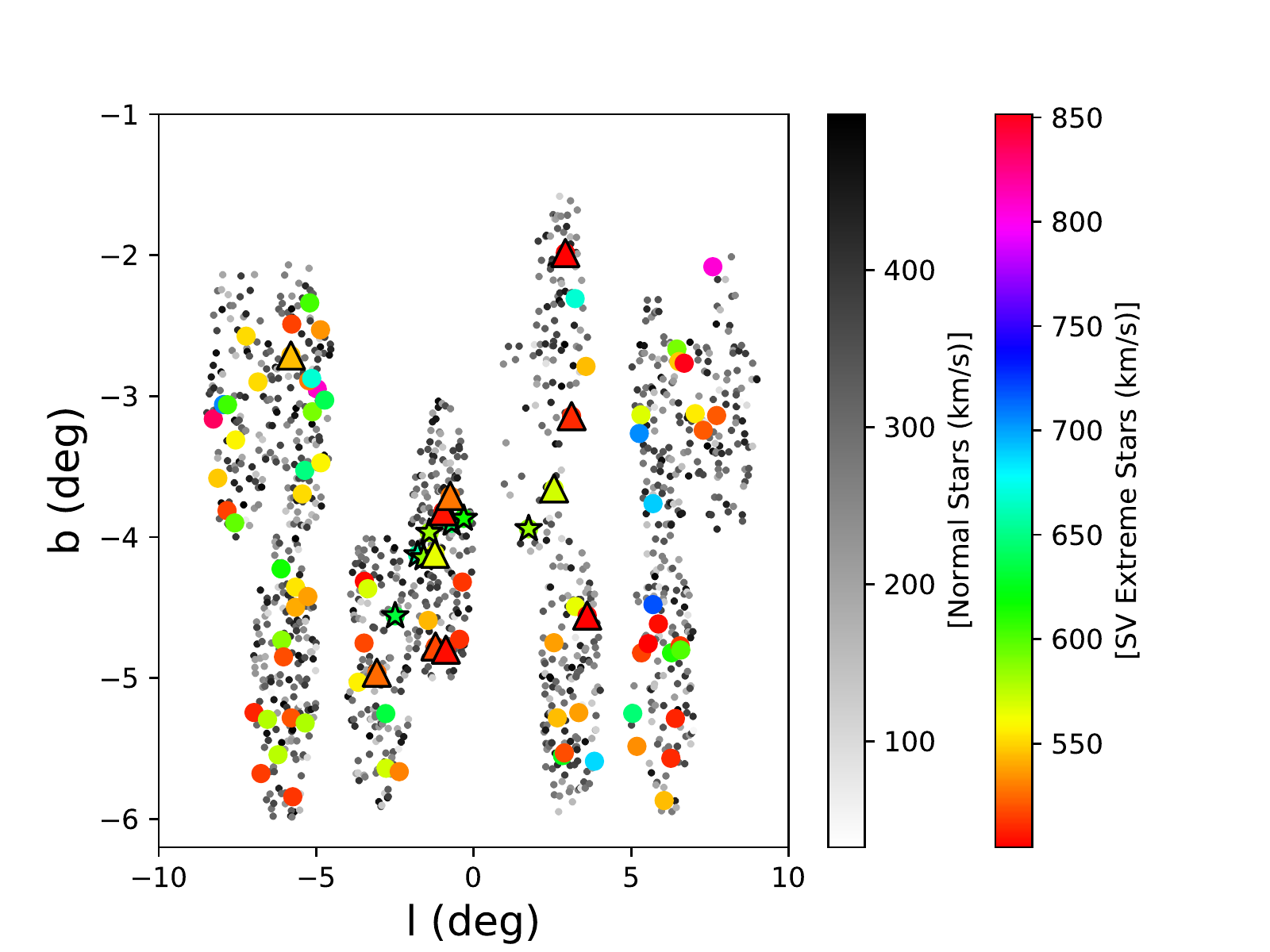}}}
\caption{The Gaia DR2 proper motions of the BRAVA-RR stars are shown.  The grey scale represents stars 
that have space velocities $<$ 500 km~s$^{-1}$ and the color scale represents stars that have space 
velocities $>$ 500 km~s$^{-1}$. The points marked with a triangle represent the stars that have orbits 
confined the to the bulge (apocenter distance $<$ 3.5~kpc). The points marked with a star represent seven stars that are similar in space velocity, latitude, and 
longitude. These points are also marked in Figure~\ref{apocenter}.
\label{motionlb}}
\end{figure*}

\begin{figure}
\centering
\mbox{\subfigure{\includegraphics[height=6.0cm]{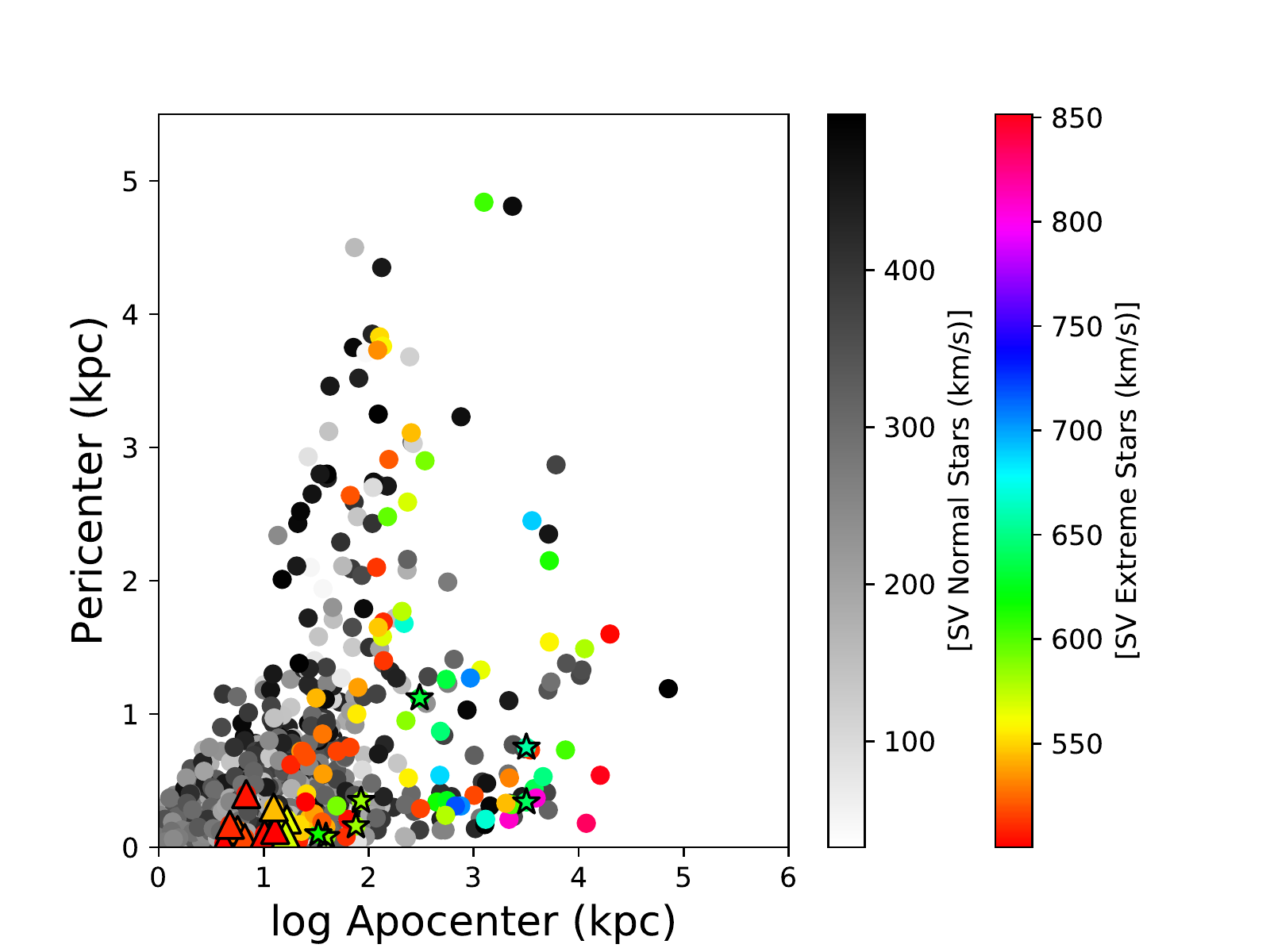}}}
\caption{The apo- and pericenters of the BRAVA-RR stars are shown.  In general, the RRLs with 
the largest space velocities are on orbits indicating they are not confined to the bulge. 
\label{apocenter}}
\end{figure}

\section{The Highest Velocity Stars}
\label{sec:emilie}
Since the discovery of hypervelocity stars \citep{brown05}, as well as the increase of several 
large spectroscopic and astrometric surveys, the exploration of stars moving with high velocities 
within our Galaxy has received growing attention \citep[e.g.,][]{boubert18, hawkins18}.  The
RRL radial velocities presented here have a larger velocity dispersion than seen in the
previously studied bulge populations (see e.g., Figure~\ref{rvcurve}) making this sample ideal for studying 
stars with large space velocities.

We turn to those RRLs with full 3D kinematic information (see \S\ref{sec:orbit}).  
The space velocity is calculated using:
\begin{equation}
SV=\sqrt{U^2 + V^2 + W^2}.
\end{equation}

We focus on those stars with space velocities greater than 2.5$\sigma$ from the mean of 
the distribution.  These encompass 7\% of the population of RRLs.  
We tentatively identify two groups of RRLs with similar space velocities and proper motions, but 
it is difficult to discern if these groups are meaningful.  One group 
of 7 RRL have space velocities of $\sim$620 km~s$^{-1}$, proper motions of 
$\mu_\alpha \sim~-$10~mas~yr$^{-1}$ and $\mu_\delta \sim -$10~mas~yr$^{-1}$), 
and also appear to be somewhat spatially aligned (see Fig.~\ref{motionlb}, right panel).  
The second group of 11 RRLs stand out because they have apocenters that indicate they may 
be confined to the bulge (i.e., apocenters less than 3.5~kpc). These stars 
have space velocities of $\sim$500 km~s$^{-1}$ and proper motions of 
$\mu_\alpha \sim~-$7~mas~yr$^{-1}$ and $\mu_\delta \sim -$10~mas~yr$^{-1}$). 
Neither of these two groups of RRLs clump in radial velocity or period, and we do not have
sufficient metallicity precision with photometric metallicities to check if they are chemically 
similar \citep[e.g.,][]{hansen16}.  

Figure~\ref{apocenter} shows the apocenters of the RRLs with high space 
velocities (see \S\ref{sec:orbit}).  Because of the wide range of apocenter distances for our sample, 
we plot the log of the apocenter distance on the x-axis.  In general, the high velocity RRLs are not confined 
to the inner 3.5~kpc of the Galaxy and are therefore likely not bulge RRLs at all, but belong to the halo.  
Follow-up spectroscopy of the high velocity stars for detailed abundances would be useful to see if any of 
these stars have a common origin. 

\section{Multiple bulge RRL populations}
\label{sec:2pops}
\subsection{Bulge Characterization}
As was shown by 
\citet{kunder15}, not all RRLs in the direction of the Galactic bulge are actually confined to the bulge.  
This is the case for other metal-poor stars located toward the Galactic bulge as well \citep[e.g.,][]{howes15}.  
The stellar density of the bulge, halo and disk populations are all considerable and overlap in the 
vicinity of the Galactic center.  Therefore, obtaining a ``pure" sample of bulge stars based on 
photometry/distance and spatial position alone, which is the usual approach when selecting 
bulge targets, is not statistically probable.
With Gaia DR2 and taking advantage of RRLs being standard candles, we are in position to use 
an orbital analysis to 
probe which RRLs are confined to the inner Galaxy, and which ones are interlopers.

Figure~\ref{rgc_division} (left panel) shows the galactocentric distances of the RRLs in our sample, 
overlaid on particles from the R1 Milky Way Galaxy simulation\footnote{http://uclandata.uclan.ac.uk/167/}, 
where a bar/bulge forms via the disk undergoing the usual buckling instability \citep{gardner14}.  
This model was designed to understand bulge formation mechanisms by tracing kinematic 
and spatial characteristics of individual particles/stars \citep[e.g.,][]{debattista05}, so is therefore 
not fully cosmological; it is a collisionless simulation with no gas, and the dark matter halo is 
a rigid potential.  This model developed a strong peanut structure \citep{debattista05} and 
can be used to explore the nature of the X-shape of the Milky Way bulge \citep[e.g.,][]{gardner14}.  
We scale the model with respect to the 
Local Standard of Rest 
and for the RRLs, we assume a Sun-Galactic 
Centre distance of 8.2~kpc \citep{blandhawthorn16}.  Stars designated as halo stars have 
apocenter distances which are greater than 3.5~kpc, and stars that belong to the bulge have 
apocenter distances which are smaller than 3.5~kpc (see \S\ref{sec:orbit}).  

It has already been shown that a break in the mean density 
distribution at a distance of $\sim$0.5 kpc from the center is evident in the 
bulge RRLs \citep{pietrukowicz12}, and this suggests there may be a difference 
between the RRLs closest to the Galactic center and those further out.
We separate the RRLs into those 
spatially closest to the center of the Galaxy (black circles, $\rm R_{GC} < 0.9~kpc)$, and 
those farther out (crosses, $\rm R_{GC} > 0.9~kpc)$.  
The division between centrally concentrated RRLs and bar/bulge RRLs 
was made in order to obtain a sample 
of RRLs that is as centrally confined as possible but with still enough stars to carry out 
a statistical analysis of spatial and kinematic properties.  Ideally, we would use 
a division at $\rm R_{GC} = 0.5$~kpc, but our BRAVA-RR sample does not have a statistical 
sample of stars with $\rm R_{GC} < 0.5$~kpc.
OGLE-IV avoids the plane of the Galaxy, so the OGLE-IV RRL sample 
contains essentially no stars with $\rm R_{GC}<0.4$~kpc, and peaks at 
$\sim$1~kpc \citep{pietrukowicz15}.

The RRLs that are more centrally concentrated ($\rm R_{GC}<0.9$~kpc) have 
different orbital properties than those that are not ($\rm R_{GC}>0.9$~kpc): the more centrally 
concentrated RRL have smaller vertical excursions, $\rm |Z|_{max}$ distances and 
have apo- and pericenter distances that indicate their orbits 
are tightly bound.  Both populations of RRLs have very few RRLs with eccentricities smaller 
than 0.5, As we show below, these two populations, both which are of the bulge
in a sense that they are confined within the central $\sim$3.5~kpc of the Milky Way,
represent distinct Galactic components that overlap in the inner Galaxy. 

Our division between bar/bulge RRLs and central/classical bulge RRLs almost certainly 
does not cleanly separate these two components; some RRLs at greater distances from the Galactic center 
could still belong to the classical bulge component.  A cleaner separation may be possible with inclusion 
of metallicity information.

\begin{figure}
\centering
\includegraphics[width=9cm]{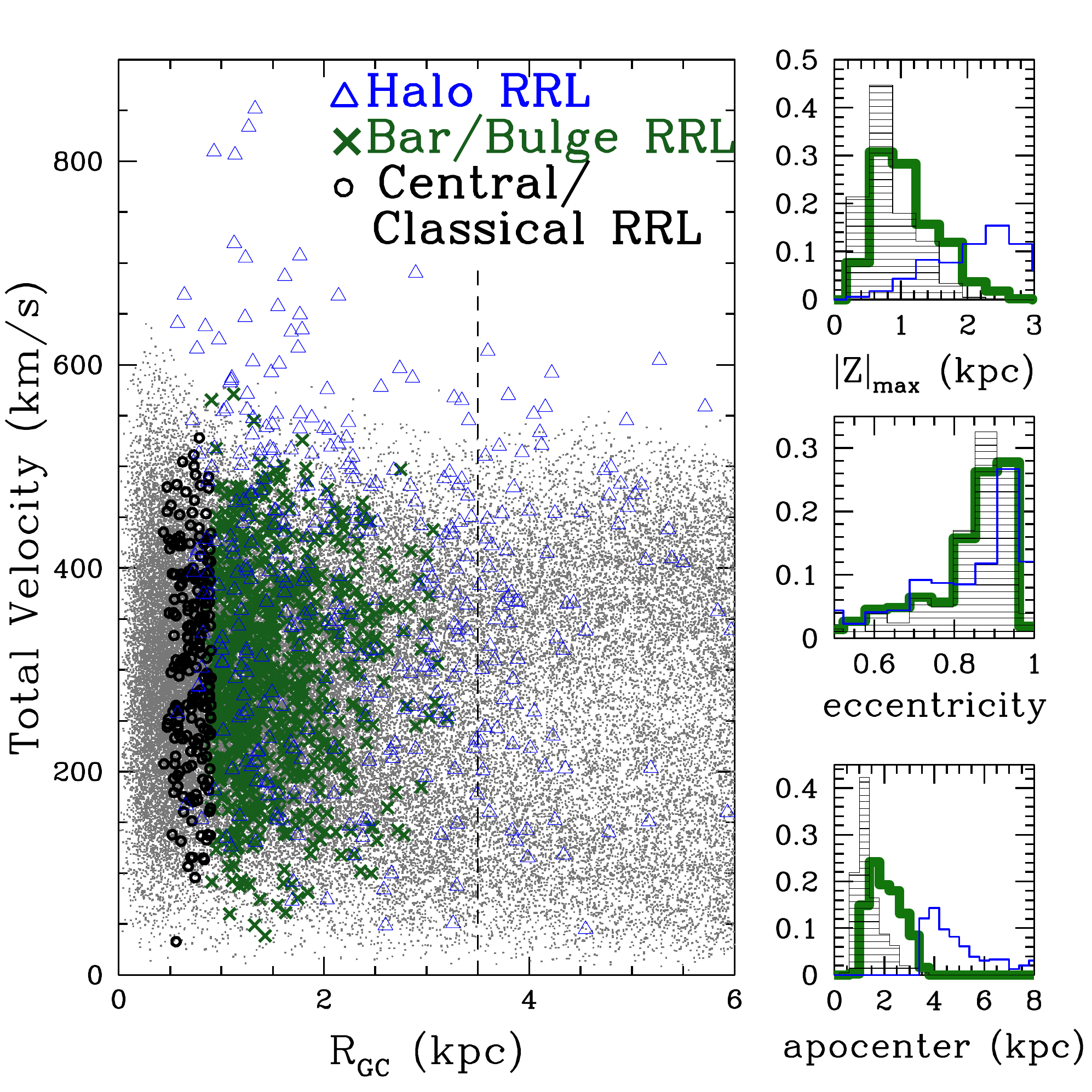}
\caption{
{\it Left:} 
The RRLs are separated into those spatially closest to the center of the 
Galaxy $(\rm R_{GC} < 0.9~kpc)$, and those farther out.  Those RRLs 
with apocenters less than 3.5~kpc are designated as bulge stars, whereas 
those RRLs with apocenters greather than 3.5~kpc are defined as halo stars.
Space velocities of stars from a Milky Way simulation (grey points) which 
includes only a disk and a halo potential \citep{gardner14} are also shown.    
The extension or the bar along the major axis of our potential is designated by 
the black dotted line at  $(\rm R_{GC} = 3.5~kpc)$.
{\it Right:} The orbital parameters of the central/classical bulge stars (black), the bar/bulge stars (green) 
and the halo interlopers (blue).
}
\label{rgc_division}
\end{figure}

\subsection{Spatial and Kinematic Distribution}
\begin{figure}
\subfigure{
\includegraphics[width=3.0in]{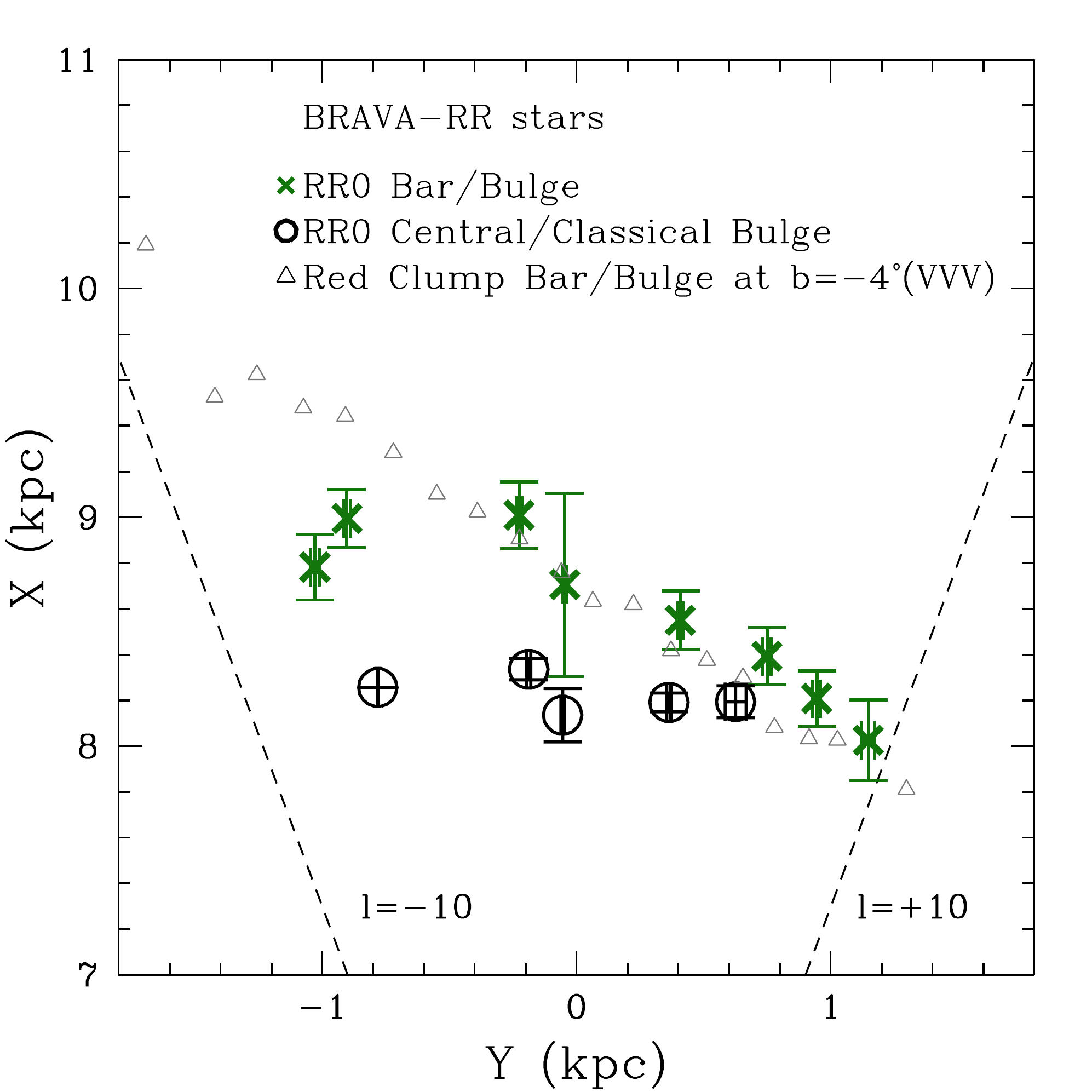} 
}
\subfigure{
\includegraphics[width=3.0in]{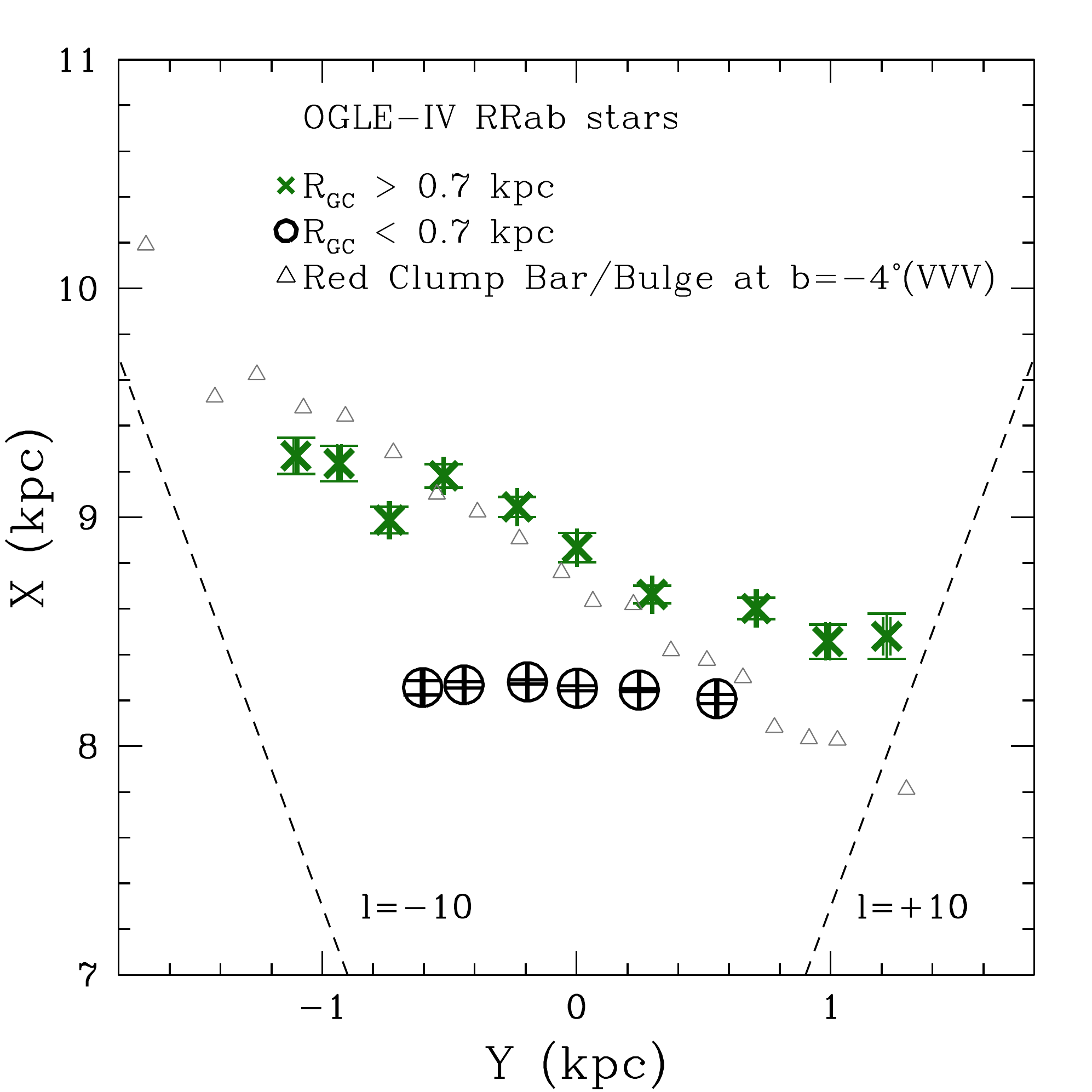} 
}
\caption{{\it Top:} The mean spatial distribution of the two bulge RRL populations 
seen in BRAVA-RR as compared to the red-clump stars from \citet{gonzalez12}.  
{\it Bottom:} The mean spatial distribution of the full OGLE-IV RRL sample split in
terms of their galactocentric distances. }
\label{spatialx}
\end{figure}

The spatial distribution of the two bulge stellar populations are shown in 
Fig.~\ref{spatialx}, and the bar as traced from VVV red clump stars 
\citep[taken from][]{gonzalez12} is also shown.  The top panel shows the spatial 
distribution of the 1389 RRLs we have space velocities for, as we can clean halo 
interlopers from this sample.  In the bottom panel we can increase the statistics by a factor of 20
by using the full OGLE-IV bulge RRL sample (although $\sim$10\% of the stars in this sample
are halo interlopers).  The results are the same--the bar/bulge RRLs trace an inclined prominent bar, 
similar to the distribution of red-clump giants in the bulge \citep[see also, e.g.,][]{stanek97, cao13}.  
In contrast, the central/classical bulge RRLs do not.  
That there are two populations of bulge RRLs, one that has a bar-shaped
spatial distribution and one that does not
could resolve disagreement for the axisymmetric RRL geometry seen using
near-infrared VVV observations\citep{dekany13}, and the view provided from optical OGLE
photometry, in which the RRLs do appear to follow the elongated spatial distribution of the bar \citep{pietrukowicz15}.

\begin{figure}
\subfigure{
\includegraphics[width=3.0in]{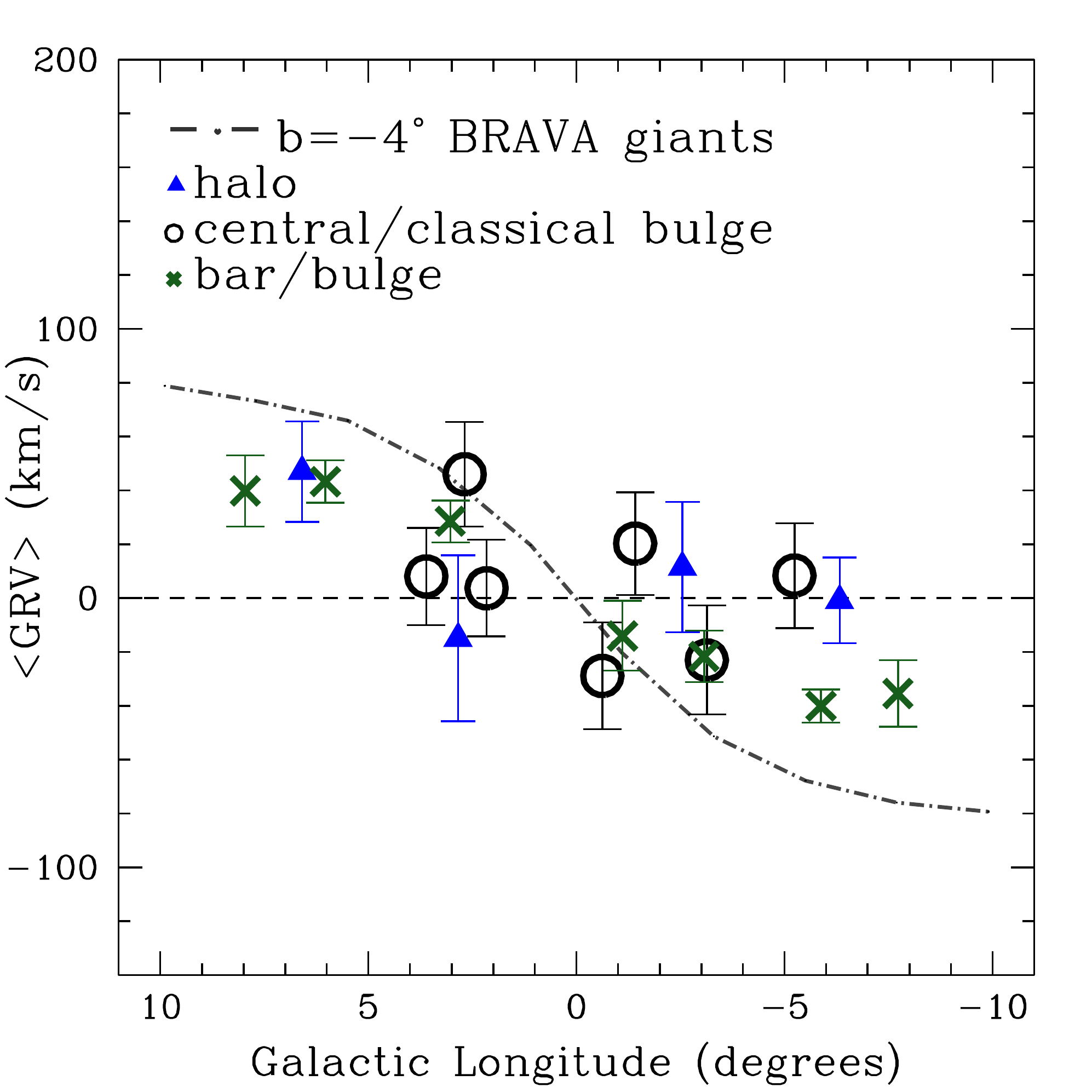} 
}
\subfigure{
\includegraphics[width=3.0in]{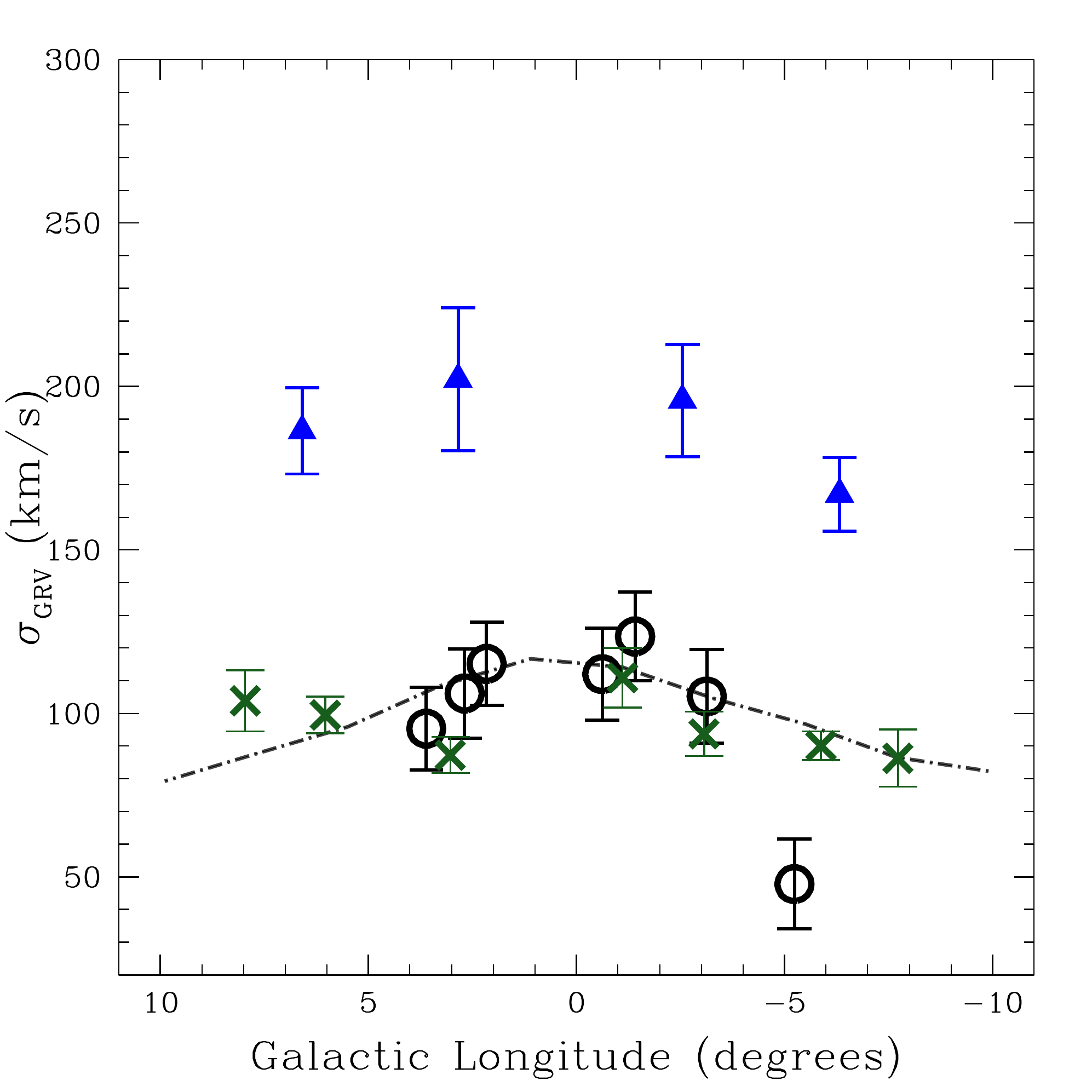} 
}
\caption{ {\it Top:} The mean galactocentric velocities
for the RRLs in the two different bulge RRL populations.
The bulge model from \citet{shen10} showing these observations are consistent with a bulge
being formed from the disk is represented by the dashed lines.
The more centrally confined RRLs have kinematics distinct from the 
bar/bulge RRLs and are a non- or slowly rotating population in the inner Galaxy.  
{\it Bottom:}  The velocity dispersion profile for the RRLs in the two different bulge RRL populations.
\label{rotcurve}}
\end{figure}
In Figure~\ref{rotcurve}, the rotation curve of the central/classical bulge RRLs and the 
bar/bulge RRLs is shown.  The bar/bulge RRLs rotate in a similar fashion to the 
giants \citep{ness13, kunder12, zoccali14, ness16}, whereas the 
central/classical bulge RRLs deviate slightly from this curve.  
The central/classical bulge rotates more similar to the halo stars passing 
through the inner Galaxy, although they have a considerably lower velocity dispersion.
The velocity dispersion of the halo found in the inner Galaxy agrees with models 
that suggest an exponentially increasing velocity dispersion with decreasing distance 
from the center of the Galaxy \citep{battaglia05}.

Both bulge RRL populations have similar velocity dispersions of 
$\sim$100~km~s$^{-1}$.  This is important when considering if 
kinematic fractionation \citep{debattista17} 
is the explanation for the two different RRL populations. 
In this scenario, all stars originate from the disk but, due to having different ages (difference of $\sim$1.5-2~Gyr) 
and initial velocity distributions, are manifest differently into the box/peanut bulge.  For example, 
kinematic fractionation could give rise to a weakly barred spatial distribution 
for a hotter and older initial population and a barred spatial distribution for a cooler 
and younger population.  We do not observe that the central RRLs have a larger velocity 
dispersion as compared to the bar/bulge RRLs, although our velocity dispersions are 
subject to the exact selection criteria of the two populations, and this is not perfect.

It is also not believed that the age spread of the inner Galaxy RRL 
population is $\sim$2~Gyr. 
This is because the horizontal branch morphology changes as a function of
metallicity and this correspondingly changes the frequency of stars that will fall on 
the instability strip.  Age differences of a few billion years produce 
dramatic changes in the morphology of the HB \citep[see e.g.,][Fig.~6]{lee92} and 
because the metallicity distribution 
of the bulge RRL is strongly peaked at $\sim$$-$1.0~dex with a dispersion 
of $\sim$0.3~dex \citep{walker91, kunder08, pietrukowicz15}, this can be reproduced 
with these RRLs having a small age spread of $\Delta t \sim$1~Gyr or less \citep{lee92, lee16}. 
An age spread larger than this would cause a wider spread of $\rm [Fe/H]$ in the bulge RRLs than 
what is currently observed.  
It may be that kinematic fractionation does occur with a smaller age range 
than $\sim$1.5-2~Gyr (Debattista, private comm.) which would then alleviate this tension.  
Further and more 
detailed simulations of kinematic fractionation may prove successful in 
reproducing the RRL results presented here.

The high velocity dispersion of the inner Galaxy RRLs first 
reported in data release 1 (DR1) of BRAVA-RR \citep{kunder16} arose from 
halo interlopers passing through the inner Galaxy, that now, with Gaia, we are 
able to isolate.  It is likely that similarly high velocity dispersions seen from samples 
of very metal-poor stars ($\rm -3 < [Fe/H] < -$1.5~dex) toward in the bulge by e.g., \citet{arentsen20} 
are also caused by a large fraction of halo interlopers that were not able to be cleaned from the sample 
do to uncertainties in finding distances to these stars.  The Gaia DR2 distances 
from parallaxes \citep[e.g.,][]{bailerjones18} have 
typical uncertainties that are $\sim$40\% or larger for stars at distances of the bulge.  Further, 
bonafide bulge stars have been shown to have parallax distances peaking at a 
distance of $\sim$5~kpc instead of at $\sim$8~kpc \citep{kunder19, arentsen20}.

One method commonly used to categorize bulges into classical and pseudobulges
is to use the $\rm V_{max}/\sigma$ parameter \citep{kormendy04}.  This can be plotted 
against $\epsilon$, the apparent flattening of the bulge, to be used to 
classify the nature of bulge systems (see Fig.~\ref{Vmaxsig}).  Pseudobulges, i.e., bulges formed via secular processes, 
have in general $\rm V_{max}/\sigma > 0.6$, whereas those that are formed via mergers, 
like classical bulges or ellipticals, have $\rm V_{max}/\sigma < 0.4$ \citep{kormendy04}. 

The BRAVA survey showed that the bar/bulge that dominates the mass of the inner Galaxy has a 
$\rm V_{max}/\sigma = 0.64 \pm 0.50$ \citep{howard08}.  Given the agreement of both the BRAVA 
radial velocities and radial velocity dispersions to the ARGOS, GIBS and APOGEE surveys, 
the uncertainty in this $\rm V_{max}/\sigma$ value is likely smaller than this.
Adopting the \citet{weiland94} minor-to-major axis ratio of $\sim$0.6 for $\epsilon$, our 
Milky Way pseudobulge is similar to that of NGC~4565, 
a well-known edge-on spiral with a peanut-shaped bulge.  This value is below the 
most rapidly rotating bars and pseudobulges due to its minor-to-major axis ratio 
(see Fig.~\ref{Vmaxsig}).  We do not attempt to find a new $\rm V_{max}/\sigma$ parameter 
using the bar/bulge RRLs presented here for the following reasons:  
(1) our BRAVA-RR sample is more than an order of magnitude smaller 
than the BRAVA sample so we don't have the statistics of other radial velocity surveys, 
(2) our selection between bar/bulge RRLs and central/classical bulge RRLs 
is almost certainly not exact, so our bar/bulge RRL sample is likely contaminated and 
(3) the ARGOS, GIBS and APOGEE surveys find almost identical radial velocities and radial 
velocity dispersions, so the $\rm V_{max}/\sigma$ from all large radial velocity surveys agree with 
that found from BRAVA. 

To probe the properties of the central/classical bulge using the $\rm V_{max}/\sigma$ parameter, 
we adopt $\sigma$=100-120~km~s$^{-1}$ for the velocity dispersion and 20-40~km~s$^{-1}$ 
for the rotation speed of the central bulge. Therefore the central bulge component has 
$\rm V_{max}/\sigma = 0.2 - 0.4 \pm 0.1$.  The minor-to-major axis ratio of the central bulge 
component has not been measured, but as is thought to be more spherical than the bar/bulge, 
$\epsilon$ is likely less than $\sim$0.4.  We have checked the Auriga 
simulations \citep[e.g.,][]{fragkoudi19, blazquezcalero20} for an indication on the minor-to-major axis 
ratio for a classical/accreted population of stars in boxy/peanut bulge.  These 
simulations do indicate that an accreted population of stars should also exist in boxy/peanut bulges, 
and that the fraction of accreted stars in the boxy/peanut region can be between 13\% to 80\% for stars 
with metallicities of $\rm [Fe/H] < -$1~dex (similar to the RRLs), and between 0.2\%-10\% for stars 
with all metallicities.  Unfortunately, a specific value of minor-to-major axis ratio for the accreted part of the bulge 
is not evident from these simulations, so the uncertainty on $\epsilon$ is substantial. 
Therefore, the $\rm V_{max}/\sigma$ of the central/classical 
bulge is in good agreement with ellipticals and a prolate stellar distribution.  A quantitative measurement of 
$\epsilon$ for the spheroidal part of the bulge would be useful.

\begin{figure}
\subfigure{
\includegraphics[width=3.5in]{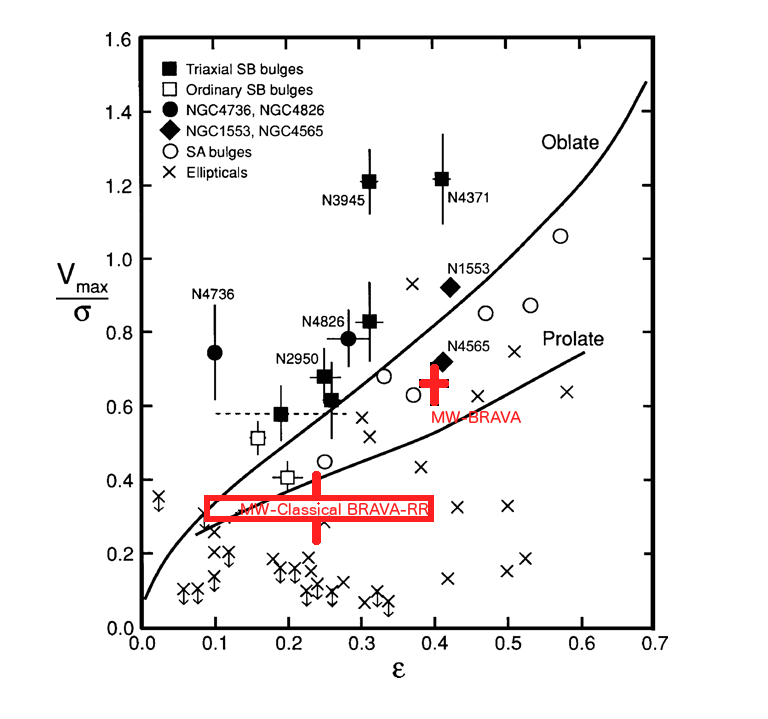} 
}
\caption{ Plot of $\rm V_{max}/\sigma$ parameter from \citet{kormendy04} with the 
approximate location of the central/classical bulge identified here indicated (open red box).  
The $\rm V_{max}/\sigma$ of the Galactic bar/bulge from \citet{howard08} is indicated by a red cross.
The $\rm V_{max}/\sigma$ is the ratio of maximum rotation velocity to mean velocity dispersion interior to 
the half-light radius and $\epsilon$ = 1 $-$axial ratio.  Oblate-spheroidal systems have isotropic velocity 
dispersions and that are flattened by rotation, and follow the ``oblate'' line (see Binney \& Tremaine 1987).  
Prolate spheroids can rotate more slowly for a given 
$\epsilon$ because they are flattened partly by velocity dispersion anisotropy; the follow the ``prolate” line. 
\label{Vmaxsig}}
\end{figure}

\subsection{Pulsation properties}
It has been shown that there are two sequences of RRLs manifest on the period-amplitude 
diagram \citep{pietrukowicz15}.  
A Kolmogorov-Smirnov (K-S) goodness-of-fit test was 
carried out to establish whether one can reject the null hypothesis that the periods of the 
inner ($\rm R_{GC} <$ 0.9~kpc) and outer (3.5~kpc $< \rm R_{GC} >$ 1.5~kpc ) RRLs come from the 
same distribution.  The K$-$S test on the full OGLE-IV sample of RRL returns a probability 
of $P=$  0.0004, below the default threshold $\rm P_{th} =$ 0.05 below 
which one rejects the null hypothesis.  Changing the distance of the outer RRLs to
$\rm R_{GC} >$ 0.9~kpc (instead of 1.5~kpc) results in K$-$S test probability of $P=$0.02.
Therefore, the pulsation properties of the central/classical RRLs appear to be different than 
the bar/bulge RRLs.  In particular, a visual inspection shows that the bar/bulge RRL population exhibits more of a 
continuum of periods, whereas the central/classical RRLs exhibits a stronger double 
sequence in the period amplitude diagram.

The double-sequence in the period-amplitude diagram was recently interpreted to indicate 
that the bulge RRLs are an accreted population that formed from 
primordial building blocks in a hierarchical merging paradigm \citep{pietrukowicz15, lee16}.  
This is because
such a double-sequence was linked to the same process of the multiple population 
phenomena seen in globular clusters \citep{romano10, tenorio15}.  So one period-amplitude 
sequence would be formed from the pristine gas in these building blocks that was 
globally enriched by supernovae.  The second sequence would be formed as the 
stars were enriched by the helium-rich winds of the first-generation of massive 
stars\citep{romano10, tenorio15}.
One test of this controversial interpretation would be to find globular cluster origin signatures, 
such as Na-O anticorrelations, in the bulge RRLs accreted in the early merger. 

There are few RRLs in bulge globular clusters (GCs), since bulge GCs are more metal-rich 
than GCs in the halo or disk, and RRLs preferentially form in metal-poor systems.  
The bulge GCs with the most RRLs are NGC~6441 and NGC~6388; these GCs 
are unusual in a sense that they not only have a large number of RRLs for their high metallicity 
\citep[each of these GCs has $\sim$30 RRL;][]{clement01}, but because the periods of these RRLs 
are longer than seen else 
where in the Galaxy \citep{pritzl01, pritzl03}.  
Very few field bulge RRLs have periods and amplitudes similar to those seen in NGC~6441 and 
NGC~6388 \citep[e.g.,][]{pietrukowicz15, kunder18b}, and it is unclear how NGC~6441 and NGC~6388 
fit into the formation scenario of the MW \citep[e.g.,][]{bellini13}.  

There are $\sim$35 RRLs in the OGLE-IV dataset in the bulge GCs NGC~6401, 
NGC~6553, NGC~6642, NGC~6304, NGC~6522, NGC~6453 and NGC~6569.  With such small numbers, 
it is not easy to come to conclusions on the similarity of the bulge GC RRLs to the field bulge RRLs based 
on periods and amplitudes alone.

Distinct period-amplitude sequences, such as seen in the bulge field RRL and the GC RRLs, 
have not been reported within the field RRLs in the MW \citep[e.g.,][]{fabrizio19}.  This is 
likely because the metallicity 
distribution of RRLs in the MW field spans a larger range in $\rm [Fe/H]$ than that 
of both the MW bulge RRL population and the MW GC RRL 
population \citep[e.g.,][]{fabrizio19}

High amplitude short period (HASP) RRLs are signatures of massive systems with rapid metal 
enrichment \citep[e.g.,][]{fiorentino15}.  The fraction of HASP RRLs, $f_{HASP}$, for the 
OGLE-IV RRL with both $\rm R_{GC} <$ 0.9~kpc and for the RRL 
with 3.5~kpc $< \rm R_{GC} >$ 0.9~kpc is $f_{HASP}$=15\%.  
This is larger than the $f_{HASP}$=11\% found for the halo RRLs 
at a distance of 5~kpc from the Galactic center \citep{belokurov18}.  Limiting ourselves to only those RRL in the 
BRAVA-RR sample, which has the advantage that halo RRLs are cleaned from the sample, 
the fraction of HASP in the central/classical RRL is $f_{HASP}$=19\%
compared to $f_{HASP}$=15\% for the bar/bulge RRL.  These high fractions suggests that 
both the inner and the outer bulge RRLs had early chemical enrichment histories, 
and were formed in a dissimilar manner to the 
surviving Galactic dwarf spheroidals (which have $f_{HASP} \sim$6\% or smaller).
Both the bar/bulge and any potential centrally located classical bulge must have been made primarily from 
progenitor galaxies larger than those that survived to become today’s dwarf spheroidals. 

\section{Discussion and Conclusions}
Within the RRL population, we find evidence for two separate components with 
distinct spatial distributions and marginally different kinematics.  
One explanation for these components is that they formed from disk buckling and 
ended up with different spatial distributions due to having slightly different ages 
and initial velocity distributions, $i.e.,$ kinematic fractionation \citep{debattista17}.
Our observations that may conflict with this explanation is that the velocity dispersions 
of our two components are 
similar, and the age spread in the bulge RRLs is thought to be $\sim$1~Gyr or less \citep{lee92}, 
whereas the age spread in kinematic fractionation is thought to be twice this.  Still, 
at early times in the formation of the Galaxy, heating is very rapid, so it may be that an
age difference less than $\sim$2~Gyr is sufficient for kinematic fractionation to give rise to two 
different spatial distributions of inner Galaxy stars.  
The observational constraints presented here on the RRL velocity dispersion and age range will be 
valuable in determining the plausibility of producing a compact bulge component from a 
pure disk simulation under the kinematic fractionation framework.

Another explanation is that the barred RRL component 
rotating in a similar manner to the red clump and red giant 
branch stars  
formed with these stars, when the disk buckled to form the 
bar/bulge (in situ formation).  The more axisymmetric component, which is more centrally 
concentrated, is then antecedent to the formation of the bar.  In this case, the more central 
RRLs trace an older, more spheroidal component that can be identified 
with an accreted component formed in the early universe.  


Our results from the RRLs are similar in some respects to those recently reported by \citet{grady19}, 
who find that the Mira population 
toward the inner Galaxy can also be separated into two components.  The young/metal-rich Miras (5-8~Gyr)
exhibit a boxy/peanut like morphology with a characteristic X-shape, whereas the old 
Miras (9-10~Gyr) are constricted radially, with little evidence for them being part of a bar-like structure.   
They interpret this to the bar/bulge having buckled $\sim$8-9~Gyr ago, and therefore
the younger Miras are part of the bar and the older Miras are not.  
From a stellar evolution view point, the age spread of the inner Galaxy RRLs is not nearly as large as that of 
the Miras.  It is striking that we see two populations of RRLs despite their 
much smaller spread of ages, and it is difficult to reconcile this with 
a bar formation and buckling taking place 8-9 Gyr ago.  This is because we see 
here ($e.g.,$~Fig~\ref{spatialx}) that a large fraction of the bulge RRLs also belong to the bar/bulge, 
and RRLs are only formed in systems older than $\sim$10~Gyr \citep{walker92, lee92}.

Although RRLs represent only $\sim$1\% of the Galactic bulge population \citep{pietrukowicz12, nataf13}
it need not mean that the classical bulge is a trace component.  
There may be other stars with low metallicities that trace these kinematics but 
that do not evolve through the RR Lyrae phase and that are difficult to disentangle
from the other stars of those metallicities in the bulge.  
With the bulge RRLs, we can sample old, metal-poor stars in an efficient manner and use 
a sample for which good distances to trace ancient populations in the bulge, but these stars may not
the only constituents of the classical bulge.
Using proper motions of $\sim$40~million predominantly red giant branch and red clump
stars in the bulge, \citet{clarke19} find no evidence for a separate, more axisymmetric, classical 
bulge component dominating in the central parts of the bulge.  This could be because the kinematic 
signature of the axisymmetric/classical bulge RRLs and the bar/bulge RRLs is more subtle than 
their spatial signatures so it went unnoticed in \citet{clarke19} 
($e.g.,$ the kinematics presented in Fig~8 is not as distinct as the spatial 
distribution presented in Fig.~7).  Alternatively, it may be that 
the axisymmetric/classical bulge is a trace component in the bulge.

At face value, our observations do not agree with the modern bulge models which include an initial 
classical bulge such as in $e.g.$,~\citet[][]{gardner14}, their models B2 and B3.  In particular, these 
models do not predict that a classical bulge would be as prominent in the central part of the Galaxy as 
what we are seeing within the bulge RRL population \citep[see also][]{pietrukowicz12}.  However, the 
RRLs trace the oldest and metal-poor stars and these models are composed of stars of all ages and 
metallicities; we do not have the ability to separate the stars in these models by ages (or metallicities).

In the cosmological hydrodynamical simulation of \citet{buck18}, in which stars are separated by ages, 
we also don't easily find agreement with the observations presented here.  The simulation by 
\citet{buck18} indicates that all stars with ages $>$10~Gyr form a roughly spheroidal and non-barred distribution 
of stars in the central few kpc of the Galaxy; only stars with ages $\sim$8~Gyr or younger contribute to the bar.  
All of the bulge RRLs must be older than 10~Gyr \citep[e.g.,][]{lee92}, but as we have shown, they are not all part of a 
spherical/unbarred bulge.  Instead, it appears that within the oldest ($>$10~Gyr) populations 
of stars in the bulge, there are at least two groups with different spatial distributions and kinematics.  

The existing model that perhaps fits our observations best is from \citet{saha16}.  These authors have 
shown that a massive classical bulge can gain some angular momentum 
from the bar. The induced rotation is small in the center, but is significant beyond $\sim$2 bulge half 
mass radii.  Because the centrally concentrated RRL do show some rotation at the larger longitudes, this could 
be interpreted as these stars belonging to a relatively massive classical bulge.  Low-mass classical bulges 
would follow the cylindrical rotation even at the center \citep{saha12}.  We do not attempt to estimate the 
mass of the classical bulge here; we leave this for future studies, that may take into account a variety of 
bulge formation models and their evolution through time, incorporating the newly available Gaia 
proper motion observations of most of the bulge stars.

\acknowledgements
AMK thanks Victor Debattista for helpful comments during the writing of this manuscript.  We thank the anonymous 
referee for suggestions that helped the clarity and quality of the paper.
We thank the Australian Astronomical Observatory, which have made these observations possible. 
The grant support provided, in part, by the M.J. Murdock Charitable Trust (NS-2017321) is acknowledged.
AK gratefully acknowledge funding by the Deutsche Forschungsgemeinschaft (DFG, German 
Research Foundation) -- Project-ID 138713538 -- SFB 881 (``The Milky Way System'', subproject A11.
 A.P.V. acknowledges FAPESP for the postdoctoral fellowship No. 2017/15893-1 and the DGAPA-PAPIIT grant IG100319. 
 
\clearpage

\begin{table}
\begin{scriptsize}
\centering
\caption{DR2 Radial velocities of BRAVA-RR stars}
\label{lcpars}
\begin{tabular}{p{0.3in}p{0.55in}p{0.80in}p{0.55in}p{0.5in}p{0.28in}p{0.58in}p{0.28in}p{0.28in}p{0.28in}p{0.28in}} \\ \hline
OGLE ID & R.A. (J2000.0) & Decl. (J2000.0) & $\rm HRV_{\phi=0.38}$ (km~s$^{-1}$) & \# Epochs & Flag & Period (d) & $(V)_{mag}$ & $(I)_{mag}$ & $I_{amp}$ & distance (kpc)  \\ 
01743 & 17 43 38.31 & $-$34 28 23.30 & \ \ 85 & 3 & 0 & 0.5314405 & 18.19 & 16.55 & 0.687 & 12.4 \\
01773 & 17 43 49.26 & $-$34 24 13.40 & $-$147 & 3 & 0 & 0.4375252 & 18.35 & 16.78 & 0.781 & 9.9 \\
01806 & 17 44 0.89 & $-$34 27 30.40 & $-$24 & 1 & 0 & 0.5615170 & 17.93 & 16.26 & 0.535 & 8.3 \\
01860 & 17 44 18.85 & $-$34 28 13.70 & \ \  62 & 1 & 0 & 0.5062415 & 18.13 & 16.71 & 0.765 & 10.7 \\
01862 & 17 44 19.49 & $-$34 19 38.60 & \ 171 & 1 & 0 & 0.4645364 & 18.72 & 17.08 & 0.761 & 11.2 \\
01873 & 17 44 22.58 & $-$34 19 6.10 & $-$61 & 3 & 0 & 0.4239042 & 18.56 & 16.84 & 0.827 & 9.2 \\
01938 & 17 44 46.95 & $-$34 23 58.00 & $-$22 & 3 & 0 & 0.5459384 & 18.58 & 16.97 & 0.573 & 11.6 \\
01983 & 17 45 1.83 & $-$34 11 27.70 & \ \  32 & 2 & 0 & 0.4759508 & 18.05 & 16.37 & 0.544 & 7.9 \\
02000 & 17 45 5.33 & $-$35 50 53.20 & \ \  12 & 2 & 0 & 0.4803552 & 17.98 & 16.53 & 0.859 & 9.8 \\
02016 & 17 45 7.56 & $-$35 21 22.90 & $-$65 & 2 & 0 & 0.5553001 & 17.35 & 15.81 & 0.472 & 7.2 \\
\hline
\end{tabular}
\end{scriptsize}
\end{table}
{}
\end{document}